\documentclass[a4,preprint,aps,pra,showkeys,] {revtex4-2}
\pdfoutput=1
\usepackage{lineno,hyperref,soul,amsmath,amssymb,color,graphicx,bm}
\usepackage[capitalise]{cleveref}

\usepackage{graphicx}
\usepackage[caption=false]{subfig}   
\usepackage{xcolor}       
\usepackage{longtable}
\usepackage{xparse}
\NewExpandableDocumentCommand\mcc{O{1}m}
{\multicolumn{#1}{l}{#2}}
\usepackage[utf8]{inputenc} 
\usepackage[english]{babel}
\usepackage{enumitem, etoolbox, tabularx, makecell, booktabs, float, nccmath}
\newcolumntype{P}[1]{>{\abovedisplayskip=\abovedisplayshortskip \belowdisplayskip=\belowdisplayshortskip}p{#1}}
\usepackage{booktabs,siunitx,makecell}
\usepackage{dcolumn}
\usepackage[colorinlistoftodos]{todonotes}
\usepackage{marginnote}
\usepackage{fancyhdr}
\usepackage{adjustbox}
\usepackage{multirow}
\usepackage[normalem]{ulem}
\usepackage{xcolor}
\usepackage{lipsum}

\makeatletter
\usepackage{miller}
\newcommand\footnoteref[1]{\protected@xdef\@thefnmark{\ref{#1}}\@footnotemark}
\makeatother
 
\reversemarginpar

\hypersetup{
	pdfnewwindow=false,      
	colorlinks=true,       
	linkcolor=red,          
	citecolor=blue,        
	filecolor=magenta,      
	urlcolor=cyan           
}
\usepackage{siunitx}
\usepackage{xfrac}
\usepackage{accents}
\numberwithin{equation}{section}
\makeatletter
\def\tagform@#1{\maketag@@@{(#1)\@@italiccorr}}
\makeatother
\let\orgautoref\autoref

\let\orgautoref\autoref

\let\orgautoref\autoref

\renewcommand{\autoref}[1]
{%
	\def\equationautorefname{Eq.}%
	\def\figureautorefname{Fig.}%
	\def\paperautorefname{paper}%
	\def\subfigureautorefname{Fig.}%
	\orgautoref{#1}%
}

\renewcommand{\autoref}[1]
{%
	\def\equationautorefname{Eq.}%
	\def\figureautorefname{Fig.}%
	\def\paperautorefname{paper}%
	\def\subfigureautorefname{Fig.}%
	\orgautoref{#1}%
}

\makeatletter
\def\tagform@#1{\maketag@@@{\ignorespaces#1\unskip\@@italiccorr}}%

\makeatother%
\usepackage{soul}
\usepackage{longtable}
\usepackage{enumitem}
\usepackage{etoolbox}
\makeatletter
\patchcmd\frontmatter@PACS@format{\addvspace{11\p@}}{}{}{}
\pretocmd\frontmatter@keys@format{\addvspace{11\p@}}{}{}
\patchcmd{\titleblock@produce}
  {\@pacs@produce\@pacs\@keywords@produce\@keywords}
  {\@keywords@produce\@keywords\@pacs@produce\@pacs}
  {}{}
\makeatother
\newlength\mylength
\usepackage[caption=false]{subfig}
\usepackage{chemformula}
\usepackage{siunitx}

\begin{document}

\title{Predicting Ti-Al Binary Phase Diagram with an Artificial Neural Network Potential}
\author{Micah Nichols}
\affiliation{Michael W. Hall School of Mechanical Engineering}
\affiliation{Center for Advanced Vehicular Systems, Mississippi State University, Mississippi State, Mississippi 39762, USA}
\author{Mashroor S. Nitol}
\affiliation{Los Alamos National Laboratory, Los Alamos, NM, 87545, USA}
\author{Saryu J. Fensin}
\affiliation{Los Alamos National Laboratory, Los Alamos, NM, 87545, USA}
\author{Christopher D. Barrett}
\affiliation{Michael W. Hall School of Mechanical Engineering}
\affiliation{Center for Advanced Vehicular Systems, Mississippi State University, Mississippi State, Mississippi 39762, USA}
\author{Doyl E. Dickel}
\email{doyl@me.msstate.edu}
\affiliation{Michael W. Hall School of Mechanical Engineering}
\affiliation{Center for Advanced Vehicular Systems, Mississippi State University, Mississippi State, Mississippi 39762, USA}

\begin{abstract}

The microstructure of the Ti-Al binary system is an area of great interest as it affects material properties and plasticity. Phase transformations induce microstructural changes; therefore, accurately modeling the phase transformations of the Ti-Al system is necessary to describe plasticity. Interatomic potentials can be a powerful tool to model how materials behave; however, existing potentials lack accuracy in certain aspects. While classical potentials like the Embedded Atom Method (EAM) and Modified Embedded Atom Method (MEAM) perform adequately for modeling dilute Al solute within Ti's $\alpha$  phase, they struggle with accurately predicting plasiticity. In particular, they struggle with stacking fault energies in intermetallics and to some extent elastic properties. This hinders their effectiveness in investigating the plastic behavior of formed intermetallics in Ti-Al alloys. Classical potentials also fail to predict the $\alpha$ to $\beta$ phase boundary. Existing machine learning (ML) potentials reproduce the properties of formed intermetallics with density functional theory (DFT) but do not examine the $\alpha$ to $\beta$ or  $\alpha$ to D0$_{19}$ phase boundaries. This work uses a rapid artificial neural network (RANN) framework to produce a neural network potential for the Ti-Al binary system. This potential is capable of reproducing the Ti-Al binary phase diagram up to 50$\%$ Al concentration. The present interatomic potential ensures stability and allows results near the accuracy of DFT. Using Monte Carlo simulations, RANN potential accurately predicts the $\alpha$ to $\beta$ and $\alpha$ to D0$_{19}$ phase transitions. The current potential also exhibits accurate elastic constants and stacking fault energies for the L1$_0$ and D0$_{19}$ phases.
\end{abstract}

\maketitle 
\newpage
\section{Introduction}
Ti-Al alloys are indispensable due to their high melting points, exceptional specific strength and stiffness, and excellent resistance to oxidation and corrosion \cite{appel2011gamma}. These properties make them suitable for load-carrying applications in aerospace, automotive, and energy industries \cite{clemens2016intermetallic, leyens2006titanium}. While Ni-based superalloys have traditionally dominated these fields, Ti-Al alloys exhibit comparable performances at temperatures ranging between 600 and \SI{800}{\celsius} \cite{dimiduk1999gamma}. The ability of Ti-Al alloys, particularly the two-phase intermetallics like L1$_{0}$ $\gamma$-\ch{TiAl} and D0$_{19}$ $\alpha_{2}$-\ch{Ti3Al}, to operate efficiently at elevated temperatures while maintaining resistance to oxidation and creep underscores their technological importance \cite{kim2014situ,palomares2017effect}. The microstructure within Ti-Al alloys changes significantly with composition. At low Al concentrations, Ti has a low temperature $\alpha$ (HCP) phase, and a high temperature $\beta$ (BCC) phase. At an Al concentration around 12$\%$, the ordered hexagonal D0$_{19}$ ($\alpha_{2}$) phase forms. These phases, in turn, affect material properties such as ductility, fracture toughness, and creep resistance. For instance, lamellar $\gamma$-TiAl, with a combination of $\gamma$ and $\alpha_{2}$ phases, exhibits superior high-temperature performance but suffers from poor room-temperature ductility due to its complex dislocation slip and twinning systems \cite{appel1998microstructure}. Studying phase evolution within Ti-Al alloys is crucial for understanding and enhancing their mechanical behavior under operational conditions. Phase transformations dictate the material's structural integrity and performance, particularly under thermal cycling, which is common in aerospace applications \cite{jepson1970effect,sato1960ms}. Understanding these transformations helps in optimizing alloy compositions and heat treatment processes to achieve desired mechanical properties.

Molecular dynamics (MD) simulations play a pivotal role in the modeling of Ti-Al alloys by utilizing atomic-scale interactions to reproduce phenomena that can be challenging to observe experimentally. These simulations provide insights into the mechanisms of plastic deformation, phase transitions, and the impact of alloying elements on the material's properties. Furthermore, MD can predict the behavior of materials under various loading and environmental conditions, thus aiding in the design of more resilient materials. Both experimental and computational methodologies have been used in the extensive study of Ti-Al alloys to interpret the complex interplay of alloy composition, microstructure, and mechanical properties \cite{asta1992first,shang2012effects,liu2007first}. Historically, experimental approaches, such as in-situ transmission electron microscopy (TEM) experiments, have provided significant insights into basic deformation mechanisms, interface strengthening behaviors, and crack nucleation and propagation in these alloys. These studies have elucidated the role of microstructural features like grain and interphase boundaries in governing plasticity and fracture behaviors \cite{ding2015features,yoo1995mechanistic}.

Computational research efforts have primarily focused on the use of density functional theory (DFT) \cite{fu1990elastic,liu2007first,sun2013nano,lee2020first} and empirical or semi-empirical interatomic potentials such as the Embedded-Atom Method (EAM) \cite{daw1984embedded,farkas1996interatomic,zope2003interatomic} and the Modified Embedded-Atom Method (MEAM) \cite{baskes1987application,baskes1989semiempirical,kim2016atomistic} to describe the microstructural features in Ti-Al alloys. DFT calculations have shed light on the non-planar character of dislocation cores and their associated Peierls stresses, offering deep insights at the atomic scale.  One major drawback in using these quantum mechanics-based methods is the computational demand required when simulating complex dislocations or large systems \cite{koizumi2006energies,karkina2012dislocation,legros1996prismatic}.

Emperical potentials, while useful in overcoming the size and time limitations of DFT, often fall short in accuracy. For instance, EAM and MEAM potentials have been tailored to capture the directional bonding essential for modeling Ti-Al alloys but have shown discrepancies in modeling the full spectrum of plasticity and fracture phenomena across different stable phases \cite{pei2021systematic}. This has led to a partial understanding that lacks the ability to comprehensively predict material behavior under varied operational conditions, particularly at high temperatures.

To address these challenges, a paradigm shift has emerged in the field of materials science, wherein machine learning (ML) techniques are leveraged to develop interatomic potentials \cite{smith2017ani,vita2024spline,morawietz2021machine,zuo2020performance,deringer2019machine,behler2016perspective,kobayashi2017neural,dickel2021lammps}. ML potentials represent a departure from conventional methods, minimizing reliance on physical or chemical intuition by interpolating mathematical functions using reference data obtained from DFT computations. This data-driven approach has revolutionized computational materials science, accelerating the discovery of new materials by accurately capturing the underlying physics of interatomic bonding. Some of the more popular ML potentials include the spectral neighbor analysis method (SNAP), Gaussian approximation potentials (GAP), spline-based neural network potentials (s-NNP),and moment tensor potentials (MTP)) \cite{thompson2015spectral,bartok2010gaussian,vita2024spline,shapeev2016moment}. For the Ti-Al system, the most prominent ML potential employs the MTP \cite{qi2023machine} formalism and exhibits superior accuracy when compared to previous ML potential \cite{seko2020machine}. Despite this accuracy, MTP does not adequately capture the transition between pure Ti and its intermetallics. While it demonstrates excellent accuracy within the intermetallic phases, it fails to accurately predict phase evolution and the effects of solutes on the plasticity of pure Ti's $\alpha$ and $\beta$ phases.

Building on the foundation laid by previous research, the current study introduces significant advancements in the molecular dynamics simulation of Ti-Al alloys through the development and implementation of the Rapid Artificial Neural Network (RANN) framework \cite{dickel2018mechanical,nitol2022machine,nitol2023hybrid,nitol2022unraveling,nitol2021artificial}. This novel approach leverages machine learning to interpolate complex potential energy surfaces from a vast dataset derived from high-fidelity DFT calculations. Unlike previous interatomic potentials, RANN is capable of modeling both the dilute and intermetallic phases of Ti-Al alloys with a level of precision comparable to DFT calculations. RANN is also effective in seamlessly transitioning between different alloy phases without a loss in accuracy. This is particularly important for Ti-Al alloys, where phase evolution plays a critical role in determining material properties. To the best of authors' knowledge, no existing interatomic potential comprehensively models both the dilute alloy and intermetallic phases of this crucial alloy at the atomistic scale. The current study bridges the gap between the modeling capabilities of dilute alloys and intermetallics of Ti-Al.  Here,  RANN is designed to be inherently flexible, enabling it to accurately predict phase behaviors, solute effects, and the thermomechanical properties of both dilute and intermetallic phases.
\section{Methodology}
\label{methods}
\subsection{First principle calculations}
A DFT database was created using version 6.3.2 of the Vienna Ab initio Simulation Package (VASP) \cite{hafner2008ab}. The simulations utilized the generalized gradient approximation (GGA) with the Perdew-Burke-Ernzerhof (PBE) \cite{perdew1996generalized} exchange-correlation functional. A densely packed Monkhorst-Pack $k$-mesh was implemented to ensure accuracy and maintained a minimum distance between neighboring $k$-points of $2\pi \times 0.01 \text{\AA}^{-1}$ in reciprocal lattice units. The electronic wave functions were expanded using a kinetic energy cutoff of 520 eV, and a Gaussian smearing of 0.2 eV smoothed the integration over the Brillouin zone.

The database contains data with up to 15\% equilibrium lattice distortion for BCC, FCC, HCP, $a15$, $\beta-Sn$, SC, and DC structures in both pure and alloyed forms of Ti and Al. It also contains up to 15\% equilibrium lattice distortions for ordered intermetallic structures such as $\gamma-$TiAl and D0$_{19}$-\ch{Ti_3Al}. For all the previously mentioned structures, additional simulations displace the atoms by random vectors with a magnitude of up to 0.5 $\text{\AA}$ as well as random independent distortions of the simulation box lengths and angles up to $\pm$5\% to simulate finte temperature behavior between 100 K and 2200 K. Stacking faults, vacancies, free surface, and amorphous configurations for Ti, Al, and Ti-Al alloys were also included. A detailed table showing all data used for training can be found in the supplemental materials \cite{supp}. 
\subsection{RANN}
RANN  uses a multilayer perceptron artificial neural network and begins with an input layer that captures a structural fingerprint, characterizeing the local atomic environment for each atom based on the MEAM formalism \cite{dickel2020neural,baskes1997determination,lee2000second}. Two types of structural fingerprints are employed -- a pair interaction fingerprint and a three body fingerprint. The pair interaction is described by:
\begin{equation}
F_n=\sum_{j\neq i}\left( \frac{r_{ij}}{r_e}\right)^ne^{-\alpha_n\frac{r_{ij}}{r_e}}f_c\left(\frac{r_c-r_{ij}}{\Delta r}\right)S_{ij}
\end{equation}
where $r_e$ is the equilibrium nearest neighbor distance, $r_c$ is a cutoff distance to determine the neighbors, $j$, of atom $i$, $n$ is an integer, $\alpha_n$ is related to the bulk modulus as used in MEAM \cite{lee2000second}, and $S_{ij}$ is an angular screening term. The three body term is described by:
\begin{equation}
\begin{aligned}
G_{m,k}=&\sum_{j,k}cos^m\theta_{jik}e^{-\beta_k\frac{r_{ij}+r_{ik}}{r_e}}f_c\left(\frac{r_c-r_{ij}}{\Delta r}\right) \\
&\times f_c\left(\frac{r_c-r_{ik}}{\Delta r}\right)S_{ij}S_{ik}
\end{aligned}
\end{equation}
where $m$ is an integer, $\theta_{jik}$ describes the angle between the vectors from atom $i$ to atom $j$ and atom $i$ to atom $k$, and $\beta_k$ determines the length scale. Each type of fingerprint uses angular screening to reduce the neighbor list, therefore improving computational efficiency. Each layer is input into a nonlinear activation function. The result of the activation function combined with weights and biases determine the values for the next layer. 10\% of the data from each sample set is reserved for validation purposes. The Levenberg-Marquardt (LM) algorithm \cite{levenberg1944method,marquardt1963algorithm}, noted for its efficiency over traditional gradient descent methods in machine learning interatomic potentials (MLIPs) \cite{artrith2016implementation}, is utilized for training. To prevent overfitting and boost the model's accuracy, a regularization term of $\lambda=1\times10^{-4}$ is added to the loss function described by:
\begin{equation}
L_{MSE}=\frac{1}{m}\sum\left(\hat{Y}-Y\right)^2+\frac{\lambda}{2m}\sum \left\Vert W\right\Vert^2_F
\end{equation}
where $\hat{Y}$ is the energy predicted by the network, $Y$ is the energy from DFT, and $m$ is the number of training points. The output layer will always consist of one node that represents the energy of a particular atom. More information on the RANN formalism can be found in the works of \citet{dickel2021lammps} and \citet{nitol2023hybrid}. The LAMMPS implementation of RANN can be found at \url{https://github.com/ranndip/ML-RANN}.
\subsection{Generalized stacking fault energies}
When validating the generalized stacking fault energies (GSFE), we focus on the basal, pyramidal I, pyramidal II, and prismatic slip systems for the $\alpha$ and D0$_{19}$ phases, the \hkl(110) plane for the $\beta$ phase, and the \hkl(111) plane for the L1$_0$ phase. For the $\alpha$ and $\beta$ phases, the GSFE was computed for pure Ti first. Then, one Ti atom was swapped to an Al atom to observe the impact solute Al has on the GSFE. The structures used to obtain the GSFE curves were created using the atomman Python package \cite{BECKER2013277,Hale_2018}. The simulation cells ranged from 20-80 atoms depending on structure and orientation. The fault plane was placed at approximately the middle of the cell. Different structures were created by moving the atoms above the fault plane in two independent directions until the original configuration was reached. The structures were then put into a LAMMPS simulation containing periodic boundary conditions in the \emph{x} and \emph{y} directions and a free surface normal to the glide plane (\emph{z} direction). The atoms were allowed to relax in the \emph{z} direction but could not move in the \emph{x} or \emph{y} directions.
\subsection{Phase Diagram}
The phase diagram simulations are split into two parts: the $\alpha$ to D0$_{19}$ transition and the $\alpha$ to $\beta$ transition. The $\alpha$ to D0$_{19}$ transition employs the semi-grand-canonical Monte Carlo (MC) \cite{sadigh2012scalable} method in LAMMPS. The technique used in this work was motivated from the work of \citet{kim2016atomistic} Pure $\alpha$-Ti was equilibrated to the desired temperature and given a chemical potential value ($\mu$). The semi-grand-canonical MC fix in LAMMPS employs the Metropolis acceptance criterion \cite{metropolis1953equation} with the addition of a chemical potential difference term, $\Delta\mu$. This fix attempts to swap a Ti atom with an Al atom 10 times for every MD timestep and accepts the swap according to the Metropolis algorithm \cite{metropolis1953equation}. More information about the acceptance probability can be found in the work of \citet{sadigh2012scalable}. The simulations consisted of 1024 atoms and were run at varying temperatures for 40 ps. The same was done with a D0$_{19}$ structure. By plotting the equilibrated atomic percentage of Al against $\Delta\mu$ for the initial $\alpha$ structure and the D0$_{19}$ structure, a hysteresis loop can be found. The final atomic percentage of Al before the $\alpha$ phase jumps to the  D0$_{19}$ phase corresponds to the phase transition point for the $\alpha$ to D0$_{19}$ transition. The final atomic percentage of Al before the D0$_{19}$ phase spontaneously becomes the $\alpha$ phase corresponds to the phase transition point of the D0$_{19}$ to $\alpha$ transition. 

The method to obtain the $\alpha$ to $\beta$ phase boundary involved several simulations and was based on the work of \citet{dickel2018mechanical}. Simulations were run for Al concentrations from 0\% to 20\% Al in 2.5\% increments. For structures containing Al, MC MD simulations were used to place solute Al in their preferred lattice sites. First, the enthalpy of the $\alpha$, $\beta$, and liquid phases must be found as a function of temperature for varying concentrations of Al. These enthalpies are used in the calculation of the relative Gibbs free energy described later. Next, simulations inducing a solid-liquid interface must be run. The following steps describe simulations for the $\alpha$-liquid interface; however, the same procedure was used for the $\beta$-liquid interface simulations. An $\alpha$ phase simulation cell composed of 27,648 atoms was heated to 1000 K over 5 ps. Half of the cell was then heated over 5 ps to induce an amorphous region. Next, a constant enthalpy calculation (nph) was applied. This allowed the cell to change size in the direction normal to the solid-liquid interface to equilibrate at zero pressure. The equilibrated temperature was considered the melting point. Once this was complete, the Gibbs-Helmholtz relation can be integrated to calculate the relative Gibbs free energy. The Gibbs-Helmholtz relation is defined by:
\begin{equation}
\left(\frac{\partial \left(\frac{G\left( T\right)}{T}\right)}{\partial T}\right)_p=-\frac{H\left( T\right)}{T^2}
\end{equation}
where the Gibbs free energy and the enthalpy of the system at some temperature, T, is given by $G(T)$ and $H(T)$, respectively. After integration, the relative Gibbs free energy, $\Delta G$, was found, and the transition temperature between the $\alpha$ and $\beta$ phases was taken to be the temperature at which $\Delta G=0$.
\section{Results}
\subsection{Potential Creation}
\begin{figure}[!htbp]
\noindent \centering
\subfloat{\includegraphics[width=.75\columnwidth]{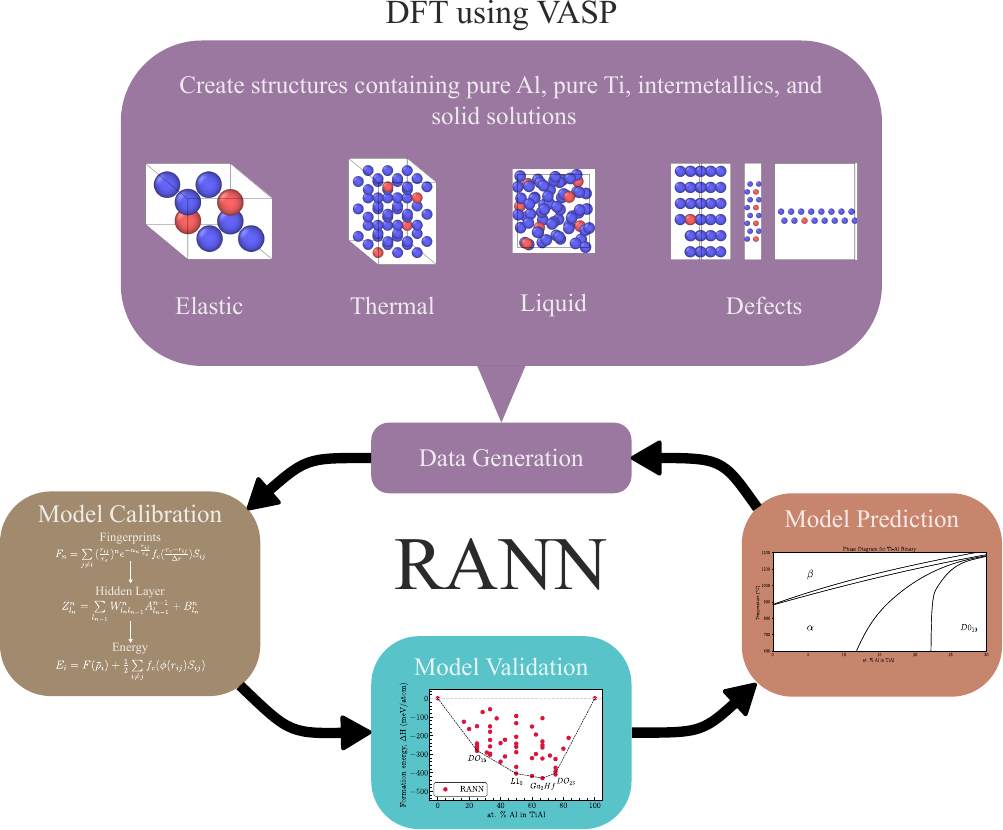}}
\caption{Summary of the RANN potential development workflow.}
\label{graphical_abstract}
\end{figure}

\autoref{graphical_abstract} illustrates the development of the RANN potential using a multilayer perceptron artificial neural network, adapted from the methods described by \citet{dickel2021lammps}. The previously mentioned DFT database used to train the potential includes a total of 62,724 simulations and 2,830,192 unique atomic environments. Each fingerprint used in the creation of the RANN potential contains metaparameters that can be adjusted to improve accuracy. Once the desired accuracy is achieved, the new potential undergoes primary validation. Primary validation includes checking the lattice parameter, elastic constants, ground-state structure, etc. against the values given by DFT to ensure that the potential has properly trained over the DFT database. If any of the tests yield incorrect results, new DFT data can be generated to inform the network of any missing information. Once the potential passes primary validation, it is tested for its capability of predicting phase boundaries. If the potential fails to predict the proper phase boundaries, new DFT data can be generated to inform the network, and the cycle begins again. The Ti-Al RANN potential can be found at \url{https://github.com/ranndip/RANN-potentials/blob/main/TiAl.nn}.

The RANN potential incorporates a single hidden layer within its network architecture, consisting of 26 neurons for the Ti network and 22 neurons for the Al network. The architecture is explicitly defined as 58$\times$26$\times$1 for Ti and 50$\times$22$\times$1 for Al, where 58 and 50 features describe the structural fingerprints for Ti and Al respectively. The metaparameters used for both networks can be found below in \autoref{rann params}.
\begin{table}[!htbp]
\caption{Metaparameters for Ti and Al.}
\label{rann params}
\resizebox{\columnwidth}{!}{%
\begin{tabular*} {1\columnwidth}{@{\extracolsep{\fill}} ccc}
\hline \hline
Fingerprint metaparameters & Ti               & Al   \\
\hline
m & $\in$ $\{$0...6$\}$ & $\in$ $\{$0...5$\}$\\
n & $\in$ $\{$-1...3$\}$ & $\in$ $\{$-1...3$\}$\\
r$_e$ $\left( \AA \right)$ & 2.924308 & 2.856975\\
$\alpha$ & 4.72 & 4.685598\\
$\beta _k$ & 1,2,5,9 & 1,2,5,9\\
r$_c$ $\left( \AA \right)$  & 8.0 & 8.0\\
$\Delta$r $\left( \AA \right)$ & 5.075692 & 5.143025\\
C$_{\emph{min}}$ & 0.49 & 0.49\\
C$_{\emph{max}}$ & 1.44 & 1.44\\
        \hline \hline
\end{tabular*}%
}
\end{table}

The network’s performance is evaluated by aggregating the outputs for all atoms in a simulation and comparing these results with total energies derived from DFT calculations. The root mean square error (RMSE) is calculated and employed as the loss function, which is minimized to refine the weights and biases. Post-training, the RMSE for the training and validation datasets are recorded at 4.68 meV/atom and 4.98 meV/atom, respectively, for the Ti-Al dataset. The high validation accuracy highlights the model's reliability and indicates that the potential is likely able to reproduce structures contained in the dataset within a 4.98 meV/atom margin of error.
\subsection{Primary Validation}
\begin{figure}[!htbp]
\noindent \centering
\subfloat[Ti]{\includegraphics[width=.5\columnwidth]{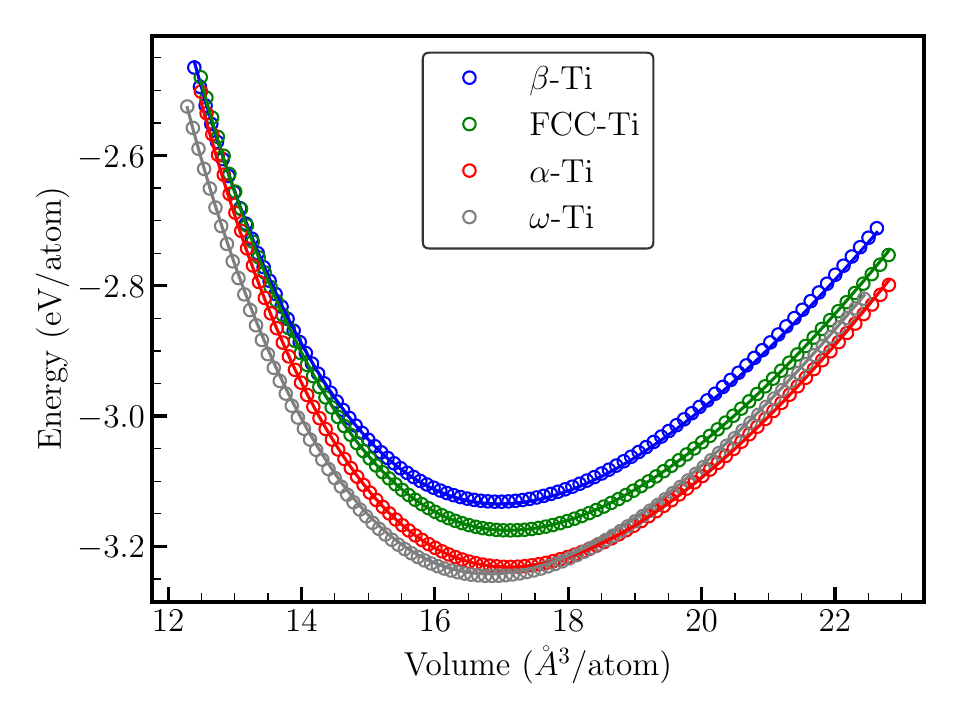}}
\subfloat[Al]{\includegraphics[width=0.5\columnwidth]{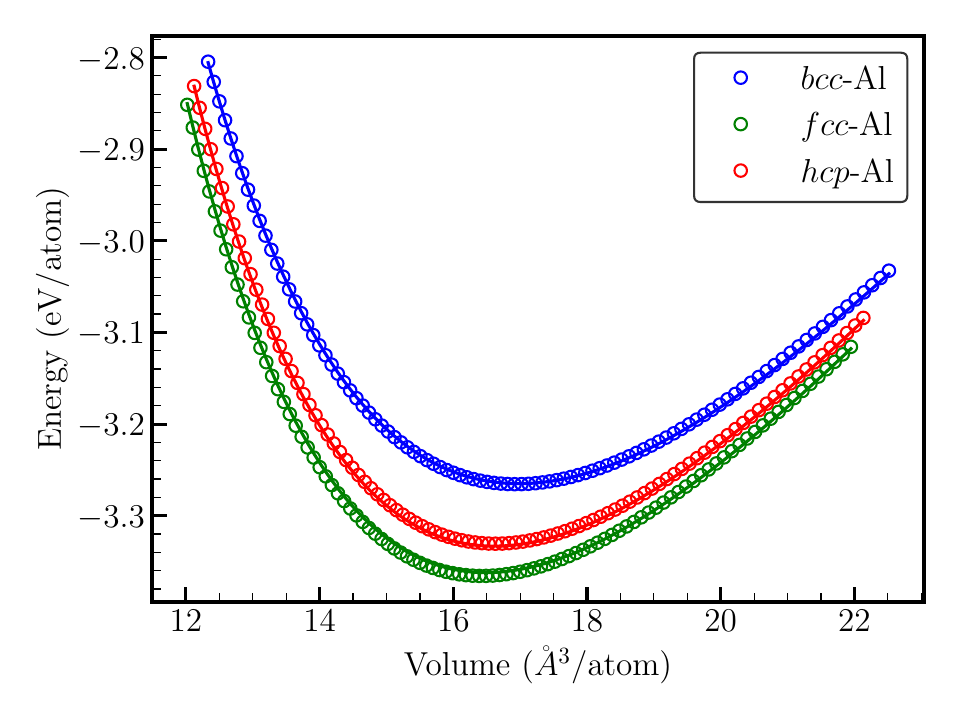}}
\caption{EV curves for (a) Ti and (b) Al. Circles represent DFT and lines represent RANN.}
\label{ev}
\end{figure}

\autoref{ev} shows an energy vs volume curve for $\alpha$, $\beta$, $\omega$, and FCC phases in pure Ti as well as FCC, BCC, and HCP phases in pure Al. RANN predicts the $\alpha$ and $\omega$ phases to be the low energy structures in pure Ti in agreement with DFT results. In pure Al, RANN follows DFT in predicting the FCC phase to be the lowest energy phase.

\begin{figure}[!htbp]
\noindent \centering
\subfloat[Relative error for RANN compared to DFT]{\includegraphics[width=\columnwidth]{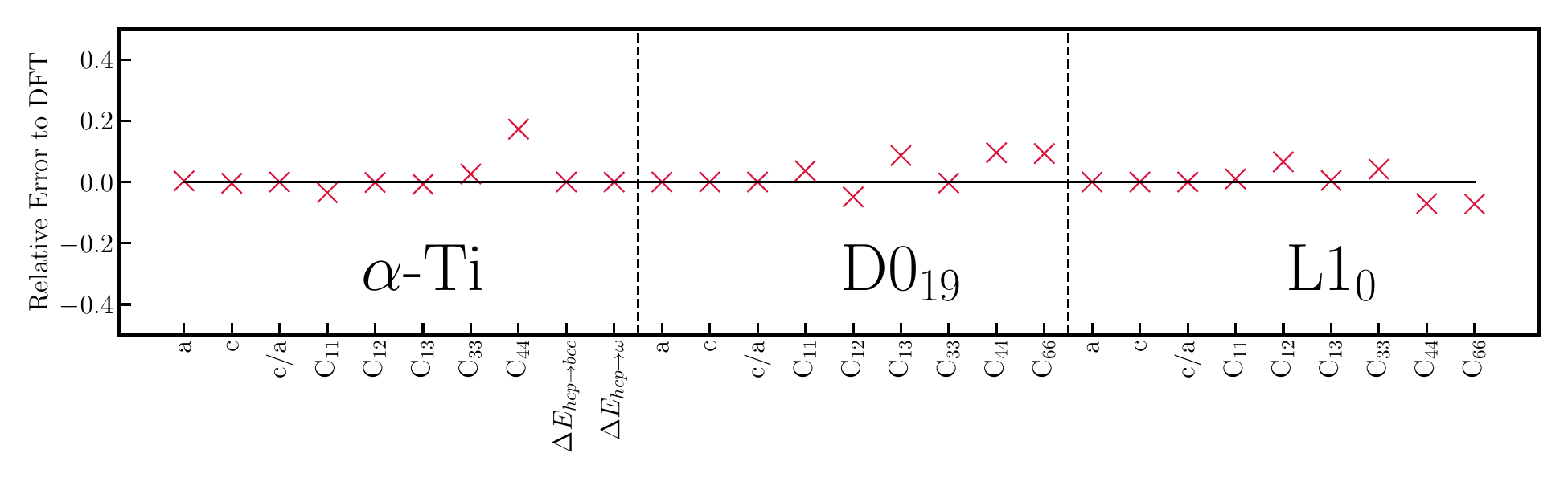}}\\
\subfloat[Convex hull for DFT]{\includegraphics[width=0.5\columnwidth]{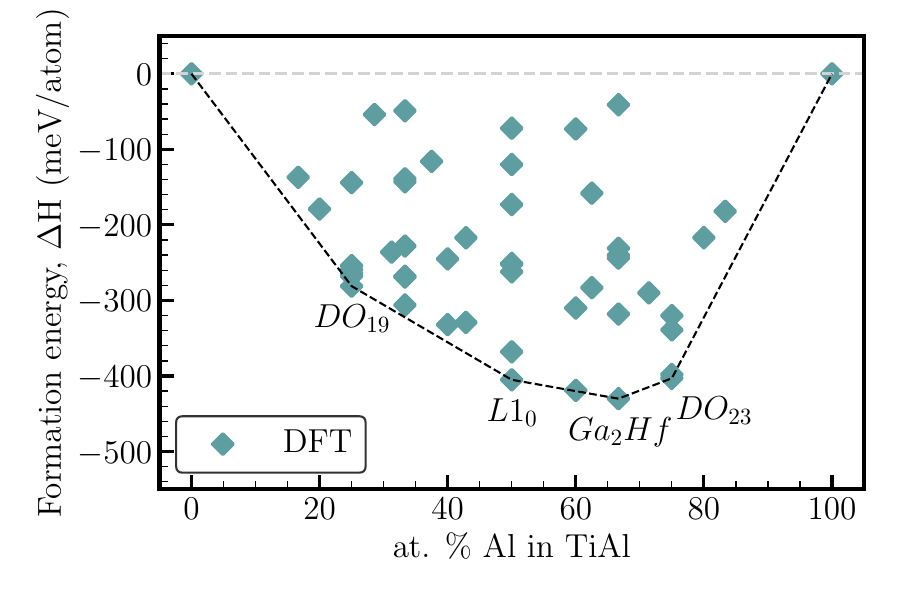}}
\subfloat[Convex hull for RANN]{\includegraphics[width=0.5\columnwidth]{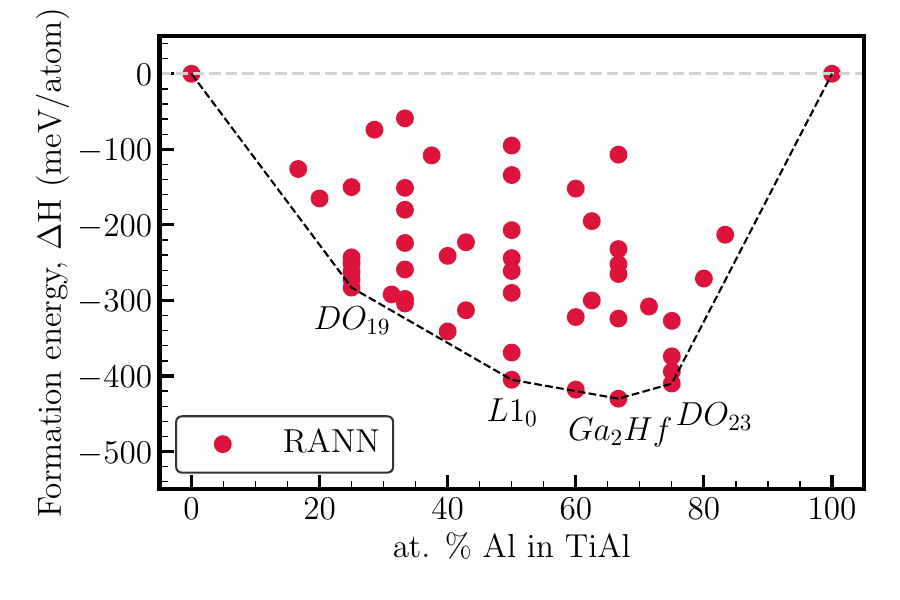}}
\caption{(a) RANN material properties for $\alpha$-Ti, D0$_{19}$, and L1$_0$ compared to DFT. (b) Convex hull for DFT and (c) RANN.}
\label{props_convex}
\end{figure}

\autoref{props_convex} shows the lattice parameters and elastic constants for pure $\alpha$-Ti, D0$_{19}$, and L1$_0$ phases as well as the convex hull for RANN compared to DFT. RANN is in good agreement with DFT calculations. The C$_{44}$ elastic constant for $\alpha$-Ti deviates the furthest from DFT at 17$\%$; however, the average deviation across these properties is 1.4$\%$. Both RANN and DFT show that the D0$_{19}$, L1$_0$, Ga$_2$Hf, and D0$_{23}$ structures are the low energy structures with increasing Al concentration. In order for the potential to accurately model the Ti-Al system, it must show that Al is soluble in Ti. DFT gives the substitutional energy of Al in Ti for the stable $\alpha$ phase to be -829 $meV$. RANN shows the substitutional energy to be -849 $meV$; therefore, RANN correctly predicts Al to be soluble in Ti. More information on the material properties for various phases can be found in the supplemental information \cite{supp} (see also references \cite{simmons1971single, yin2017comprehensive, nitol2022machine, kittel2005introduction, fisher1964single, boer1988cohesion, tyson1977surface, blakemore1985solid, kittel2005introduction, malica2020quasi, li1998ab, wang2009ab, golesorkhtabar2013elastic, shang2010first, lu2000generalized, wu2010ab, smallman1970stacking, woodward2008prediction, crampin1990calculation, denteneer1991defect, hammer1992stacking, hehenkamp1994absolute, tanaka1996elastic, fu1990elastic, liu2007first, wei2010site, zhang2017first,pearson2013handbook, pei2021systematic, tanaka1996single, fu1990elastic, liu2007first, fu2010ab, zhang2017first} therein).

\begin{figure}[!htbp]
\noindent \centering
\subfloat[$\alpha$-Ti]{\includegraphics[width=0.33\columnwidth]{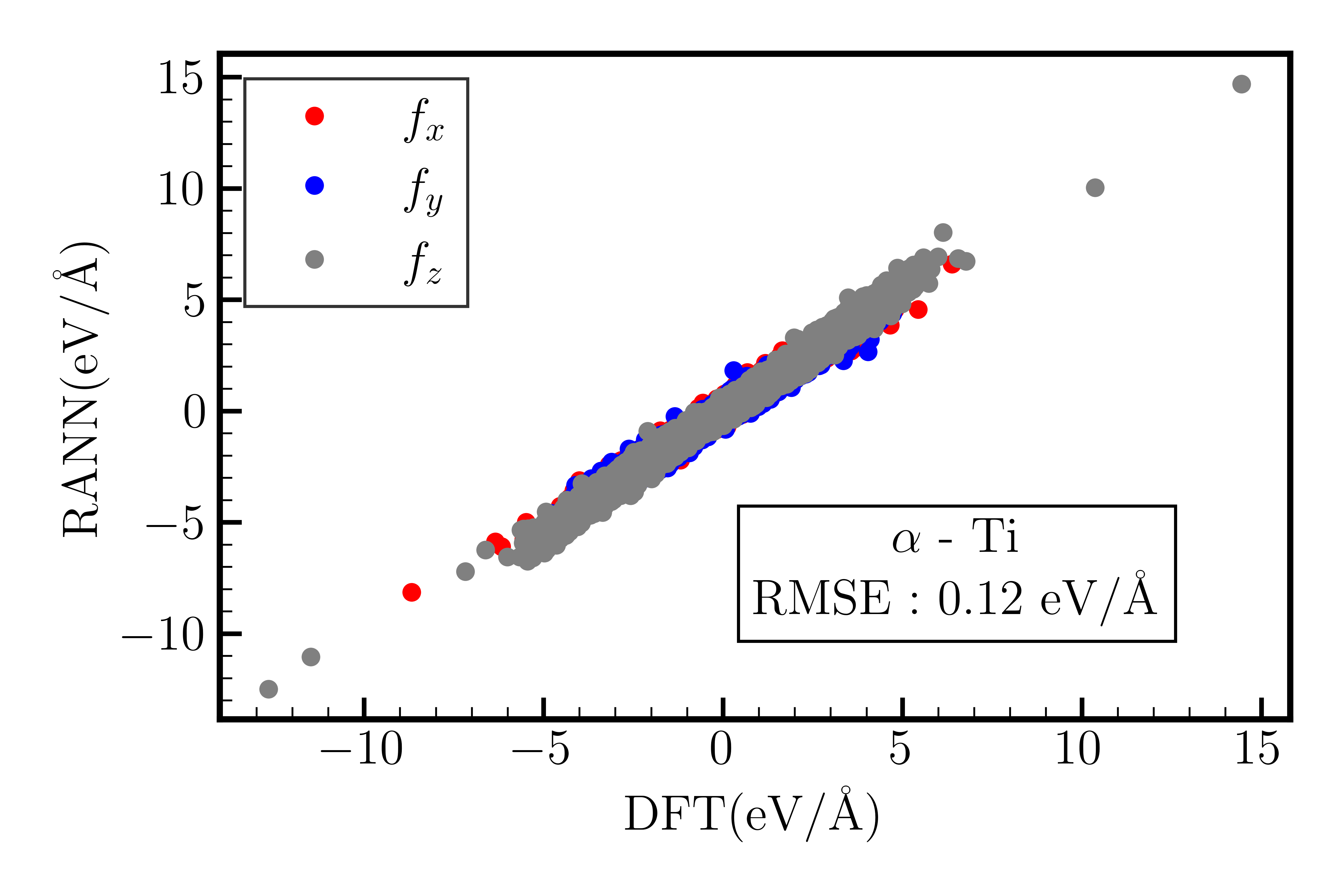}}
\subfloat[$\beta$-Ti]{\includegraphics[width=0.33\columnwidth]{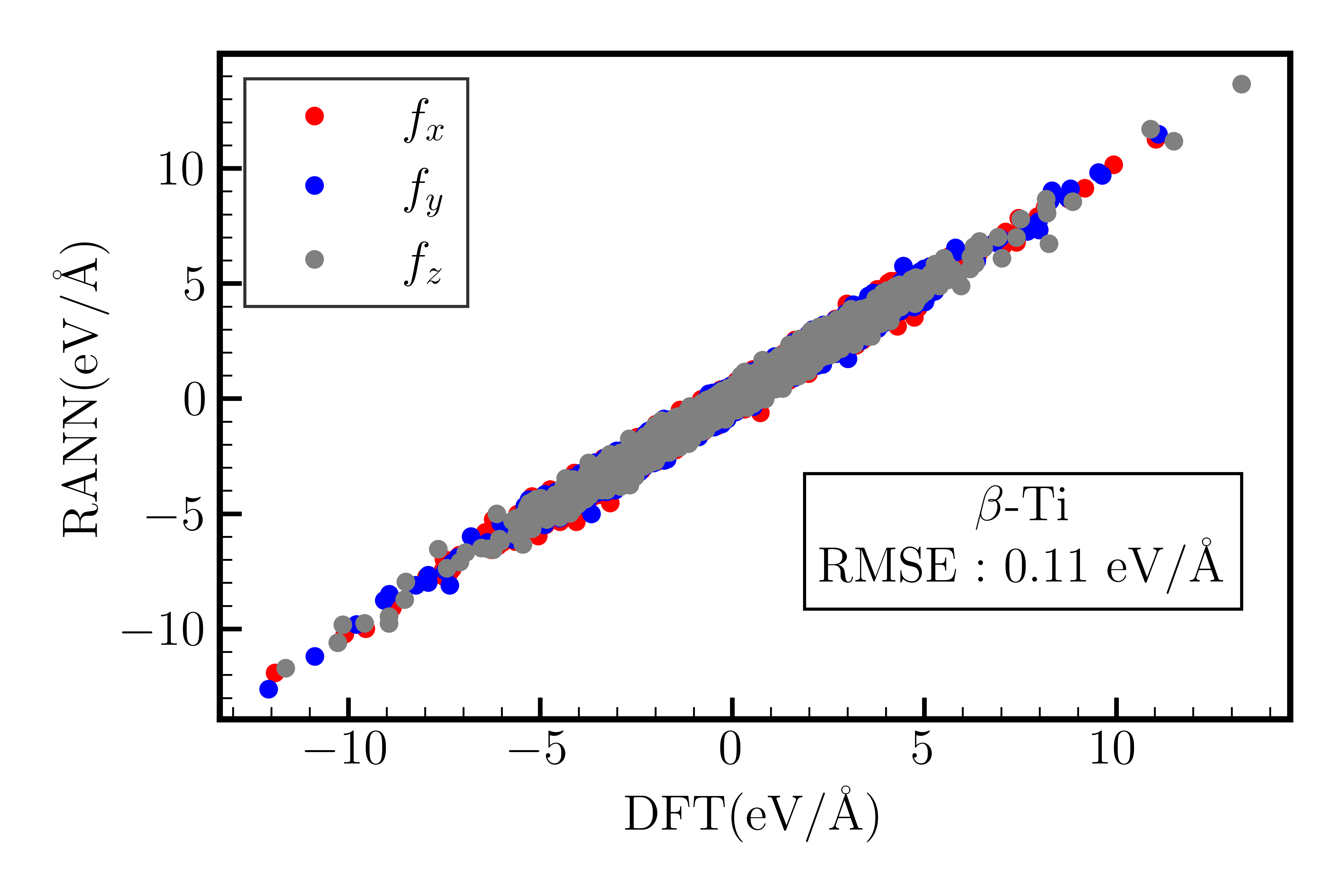}}
\subfloat[$\omega$-Ti]{\includegraphics[width=0.33\columnwidth]{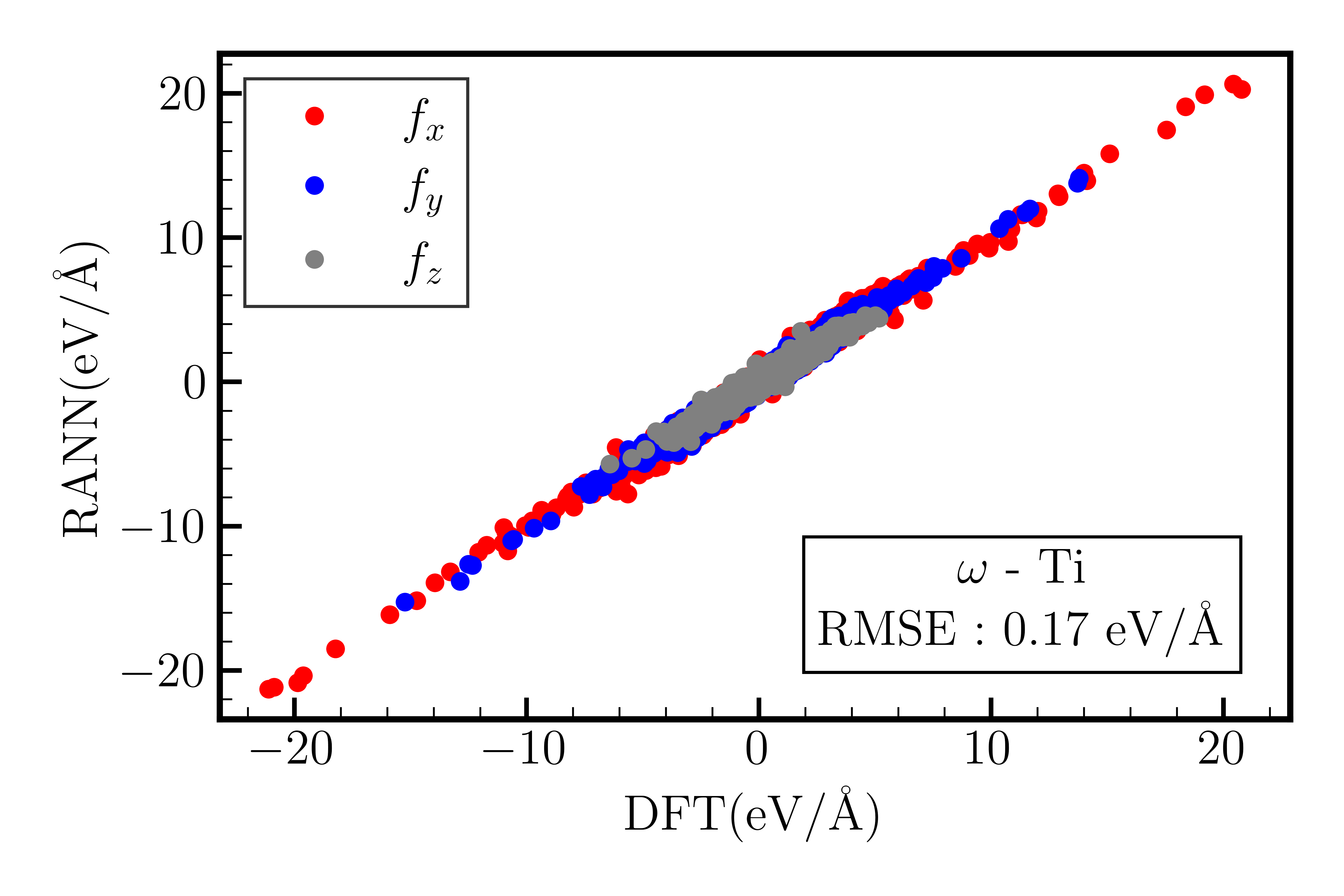}}
\caption{Force comparison for (a) $\alpha$-Ti, (b) $\beta$-Ti, and (c) $\omega$-Ti.}
\label{Ti force}
\end{figure}

Predicting the forces acting on individual atoms can be an indicator on how well an atomic potential will succeed in accurately displaying results for dynamic simulations. Here, the \emph{x}, \emph{y}, and \emph{z} components of the force acting on individual atoms for finite temperature perturbations of $\alpha$-Ti, $\beta$-Ti, and $\omega$-Ti structures are computed using DFT as well as with the RANN potential. While possible to train potentials over forces and energies from DFT, the RANN potential created in this work only trains over atomic energies. Shown in \autoref{Ti force}, RANN shows a RMSE of 0.12 eV$/$ $\AA$, 0.11 eV$/$ $\AA$, and 0.17 eV$/$ $\AA$ for $\alpha$-Ti, $\beta$-Ti, and $\omega$-Ti, respectively.
\subsection{GSFE Validation}
The presence of solute atoms along glide planes can significantly influence dislocation mechanisms. This is vital because dislocations are the main vectors of plastic deformation in metals. By understanding the interaction between solutes and dislocations, we can better predict and manage the mechanical properties of alloys. Previous studies, such as those using generalized stacking fault energy (GSFE) calculations, have demonstrated how these interactions can enhance properties like strength, ductility, and toughness \cite{kwasniak2016solid}. For instance, specific solute atoms may obstruct dislocation movement, contributing to solid solution strengthening. In Ti-Al alloys, Al solutes disrupt the emission of prismatic dislocations and alter the energy of basal/prismatic faults. This effect helps to stabilize the immobile, non-planar core of screw 1/3\hkl<11-20> dislocations, significantly enhancing the strength of $\alpha$-Ti \cite{rao2013molecular}. Understanding GSFE and the influence of solute positioning allows for the tailoring of alloy compositions to achieve desired mechanical traits, crucial for designing alloys for specific applications requiring high strength or improved ductility.

\begin{figure}[!htbp]
\noindent \centering
\subfloat[$\alpha$-Ti Basal]{\includegraphics[width=0.33\columnwidth]{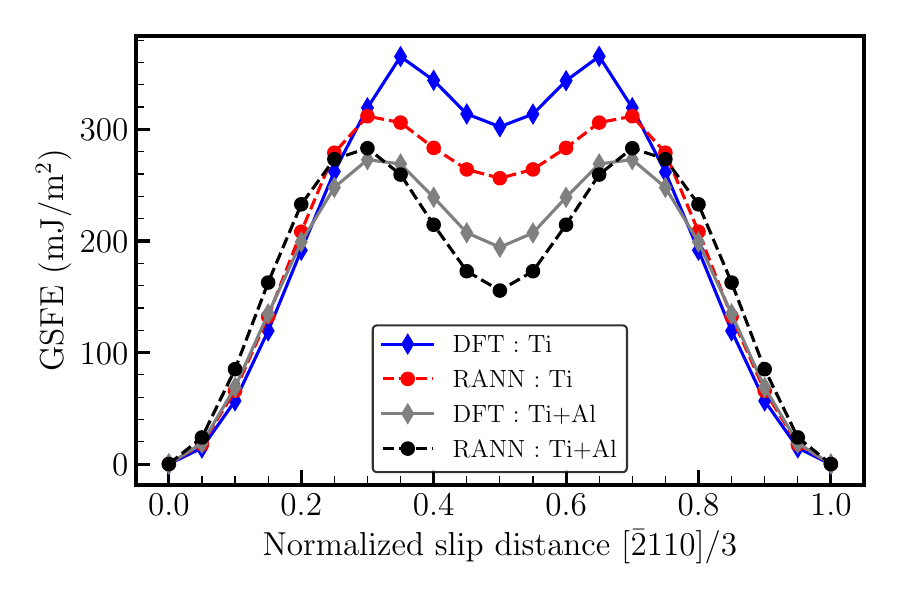}}
\subfloat[$\alpha$-Ti Basal]{\includegraphics[width=0.33\columnwidth]{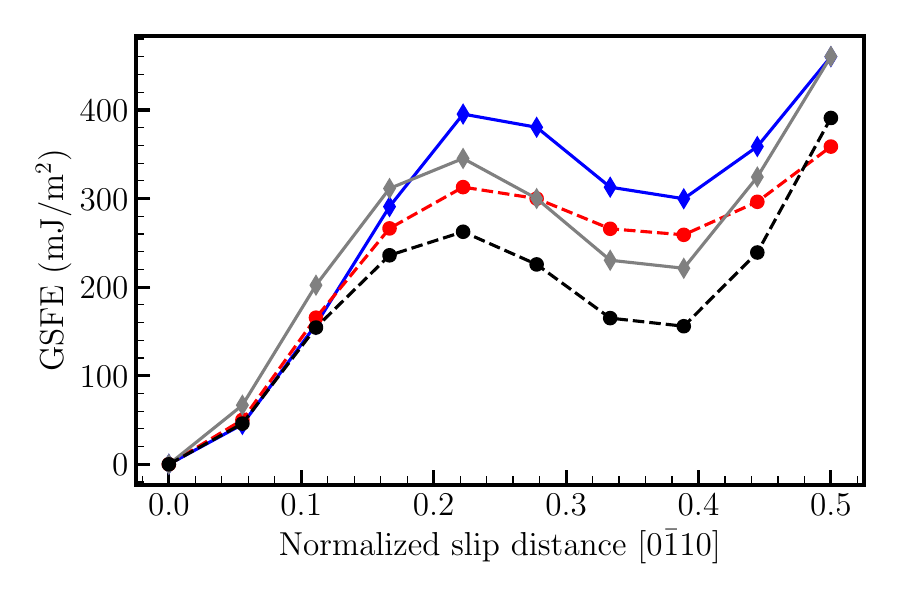}}
\subfloat[$\alpha$-Ti Prismatic]{\includegraphics[width=0.33\columnwidth]{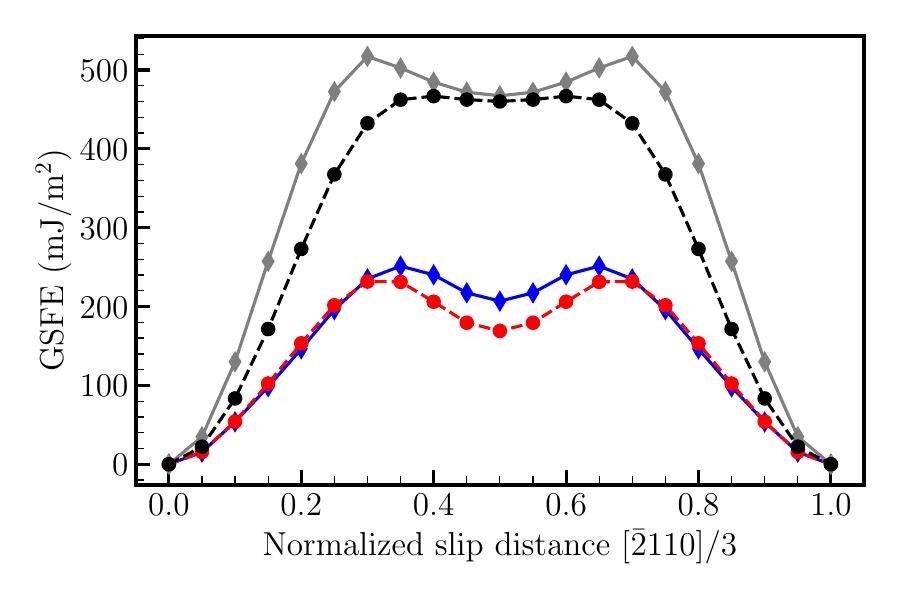}}\\
\subfloat[$\alpha$-Ti Pyramidal I]{\includegraphics[width=0.33\columnwidth]{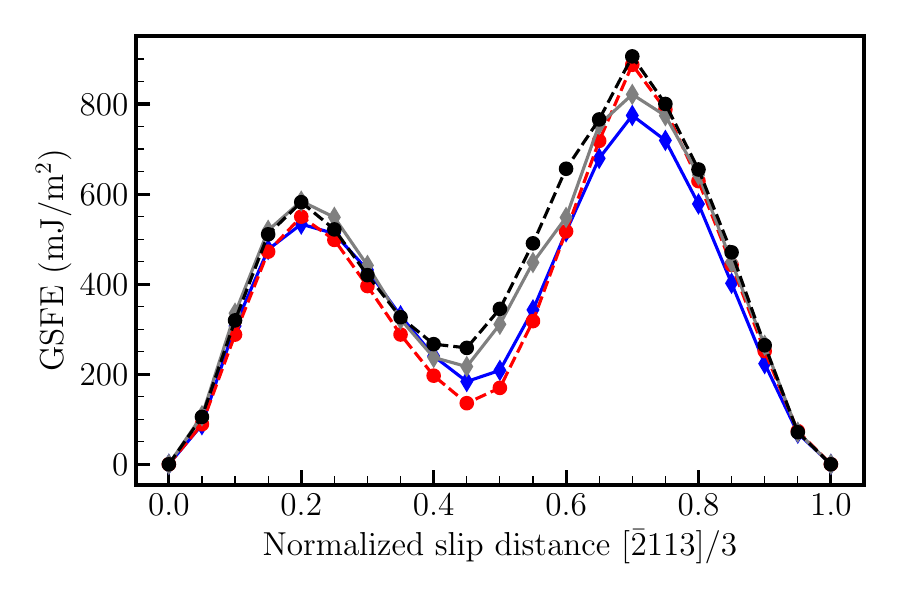}}
\subfloat[$\alpha$-Ti Pyramidal II]{\includegraphics[width=0.33\columnwidth]{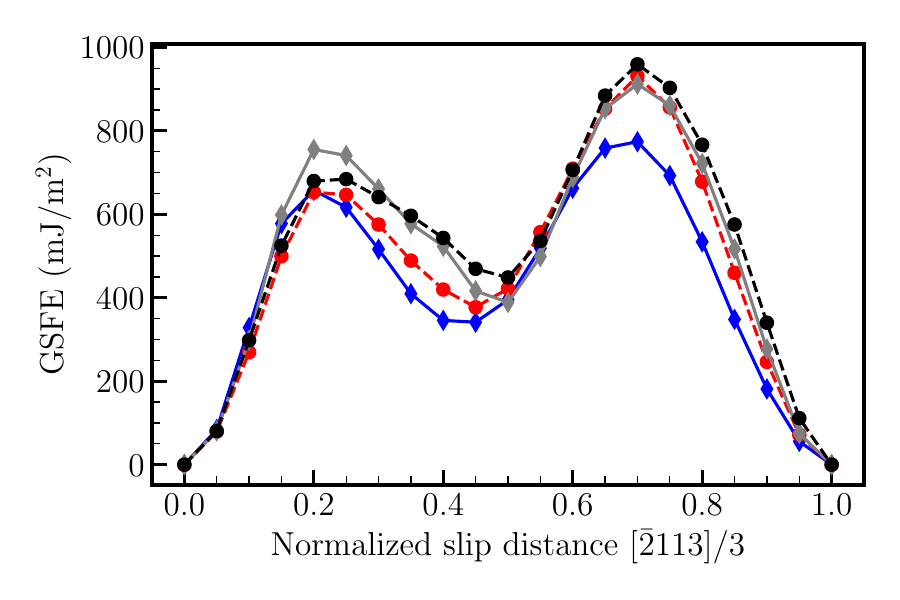}}
\subfloat[$\beta$-Ti \hkl(110)]{\includegraphics[width=0.33\columnwidth]{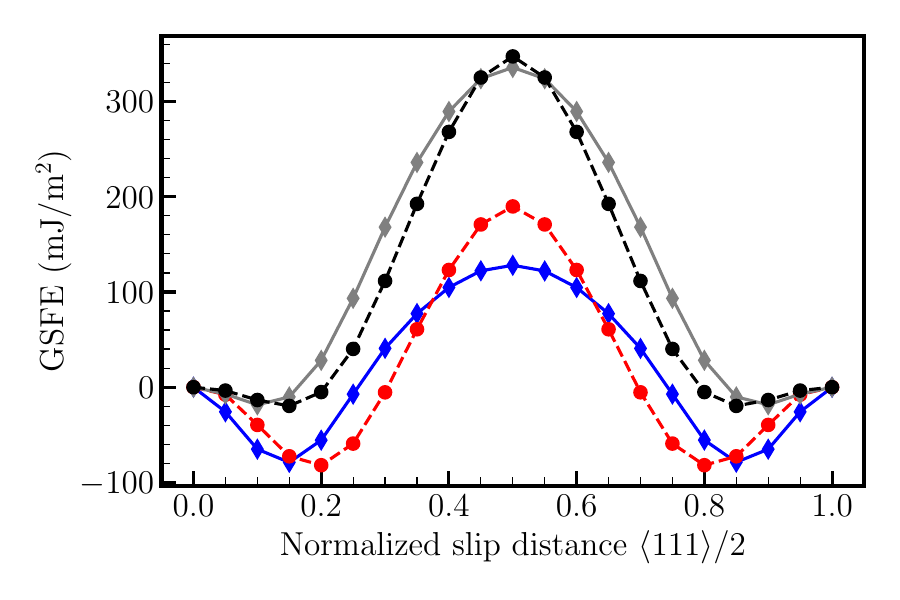}}
\caption{GSFE of Ti-Al alloy, where solute Al's position is in first nearest positions relative to the glide plane for the (a-b) Basal, (c) Prismatic, (d) Pyramidal I, (e) Pyramidal II, and (f) \hkl(110) planes.}
\label{TiAl gsfe}
\end{figure}

RANN is in good agreement with DFT for both the pure Ti case as well as when solute Al is adjacent to the glide plane. RANN shows softening in the basal plane when Al is near the glide plane as expected. Both RANN and DFT results also show a stiffening in the prismatic plane after an Al atom is placed adjacent to the glide plane. \autoref{TiAl gsfe} shows the GSFE for the mentioned planes.

\begin{figure}[!htbp]
\noindent \centering
\subfloat[Basal]{\includegraphics[width=0.5\columnwidth]{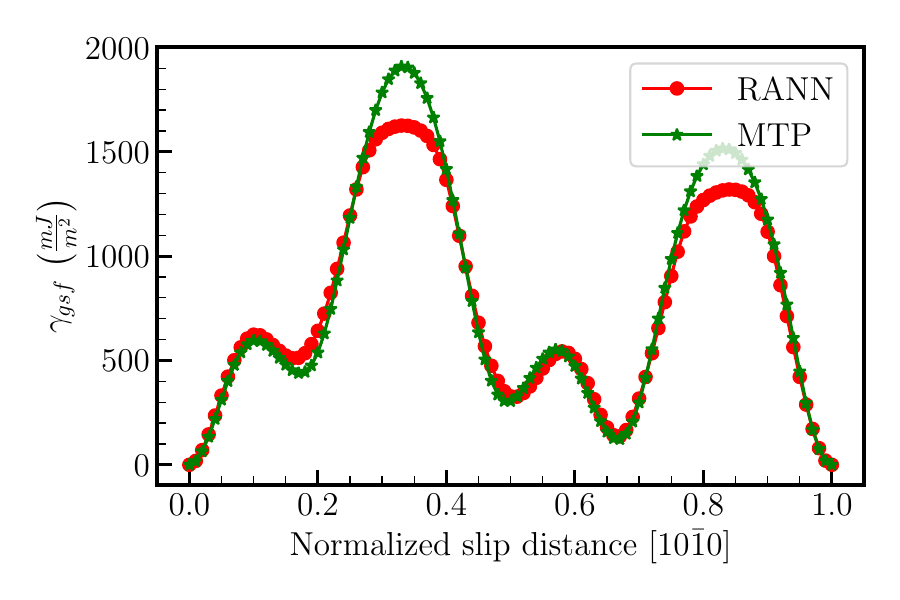}}
\subfloat[Pyramidal I]{\includegraphics[width=0.5\columnwidth]{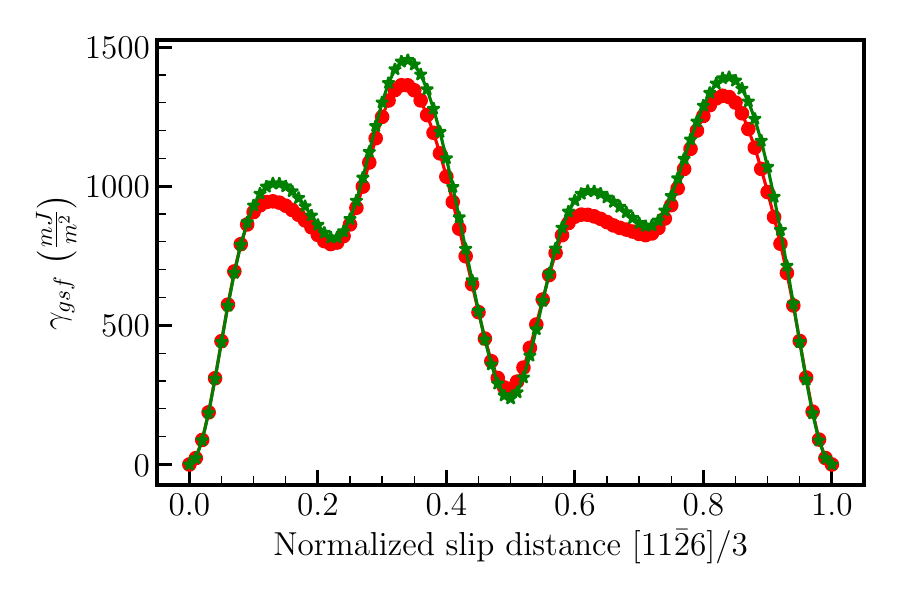}}\\
\subfloat[Pyramidal II]{\includegraphics[width=0.5\columnwidth]{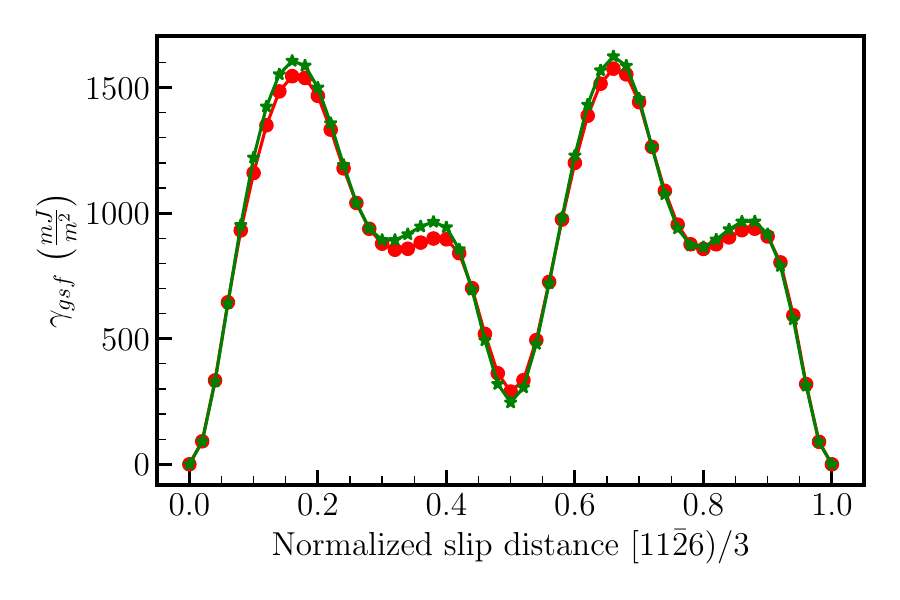}}
\subfloat[L1$_0$ \hkl(111)]{\includegraphics[width=0.5\columnwidth]{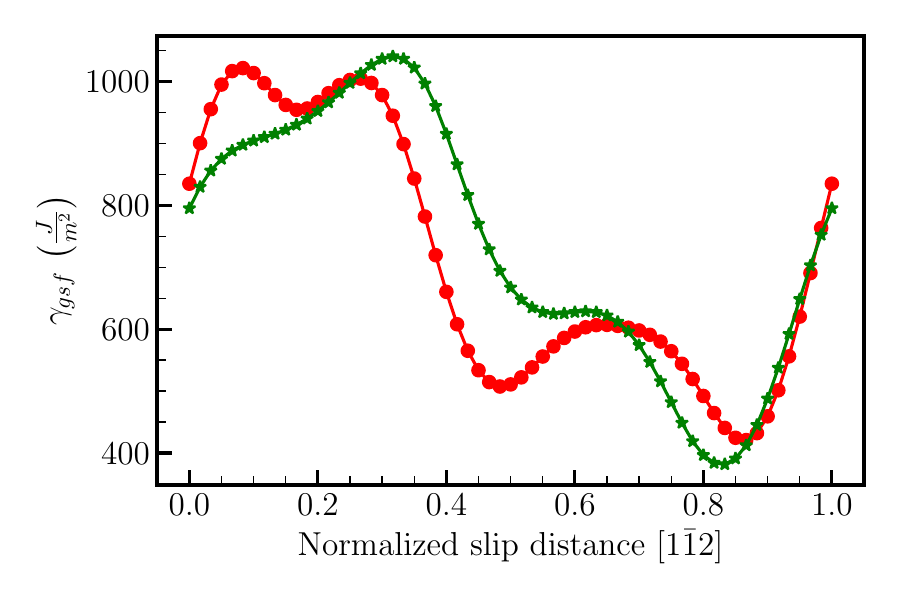}}\\
\subfloat[Prismatic Narrow]{\includegraphics[width=0.33\columnwidth]{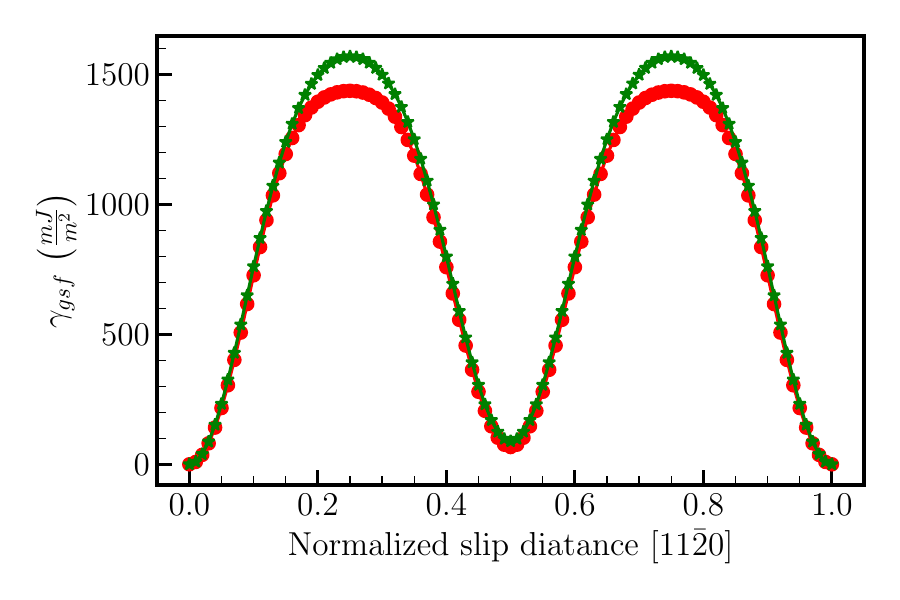}}
\subfloat[Prismatic Wide I]{\includegraphics[width=0.33\columnwidth]{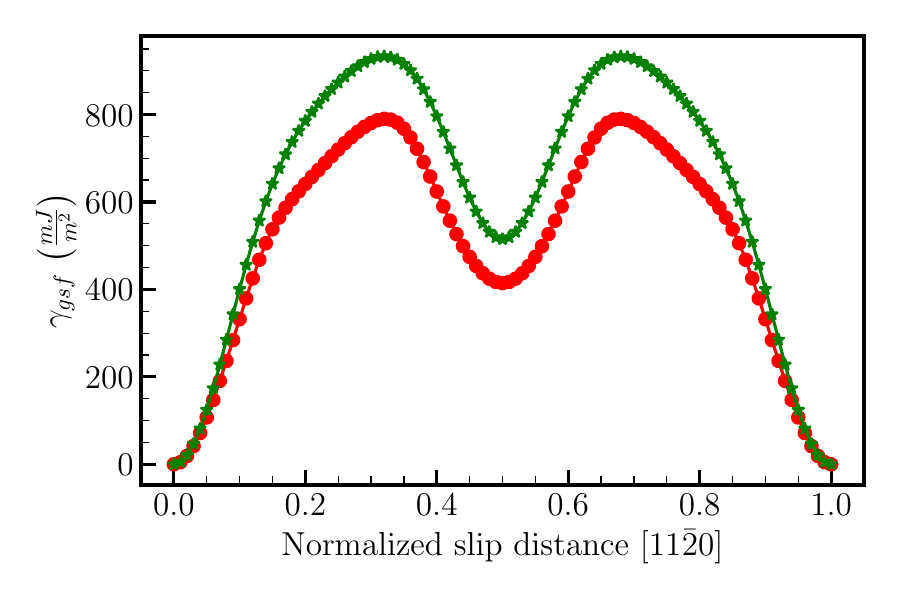}}
\subfloat[Prismatic Wide II]{\includegraphics[width=0.33\columnwidth]{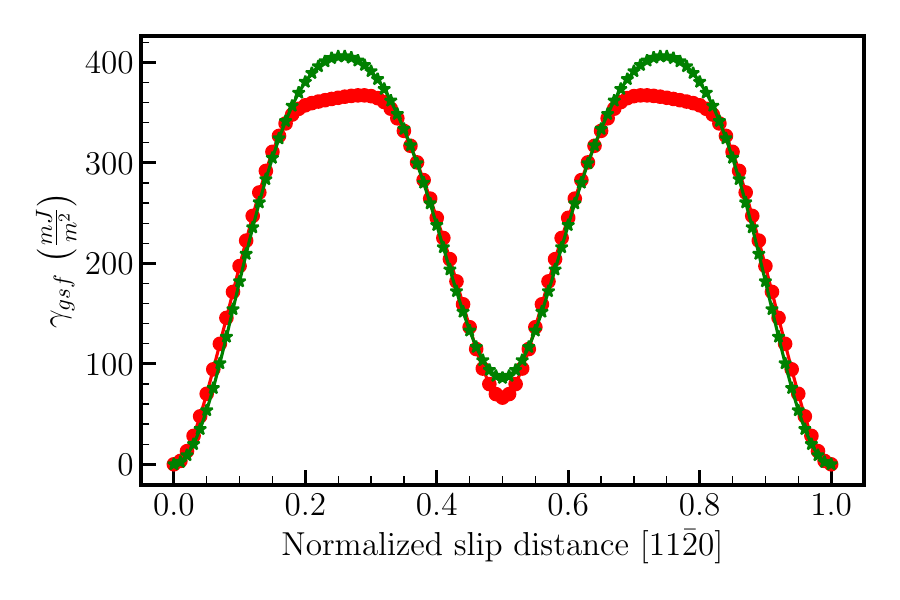}}
\caption{GSFE for (a-c,e-g) D0$_{19}$ (a) Basal, (b) Pyramidal I, (c) Pyramidal II, (d) L1$_0$ \hkl(111), (e) Prismatic Narrow, (f) Prismatic Wide I, (g) Prismatic Wide II planes.}
\label{gsfe_formed}
\end{figure}

The GSFE was also calculated for several planes in the formed intermetallic phases. The basal, pyramidal I, pyramidal II, prismatic narrow, prismatic wide I, and prismatic wide II planes were examined for the D0$_{19}$ phase, and the \hkl(111) plane was examined for the L1$_{0}$ phase. MTP is very accurate in predicting the GSFE for intermetallics \cite{qi2023machine}. Shown in \autoref{gsfe_formed}, Rann is found to be in agreement with the the GSFE predicted by MTP for intermetallic structures. A comparison of the GSFE with one solute Al atom between DFT, MTP \cite{qi2023machine}, and MEAM \cite{kim2016atomistic} can be found in the supplemental information \cite{supp}.
\subsection{Potential Prediction}
\begin{figure}[!htbp]
\noindent \centering
\subfloat{\includegraphics[width=1\columnwidth]{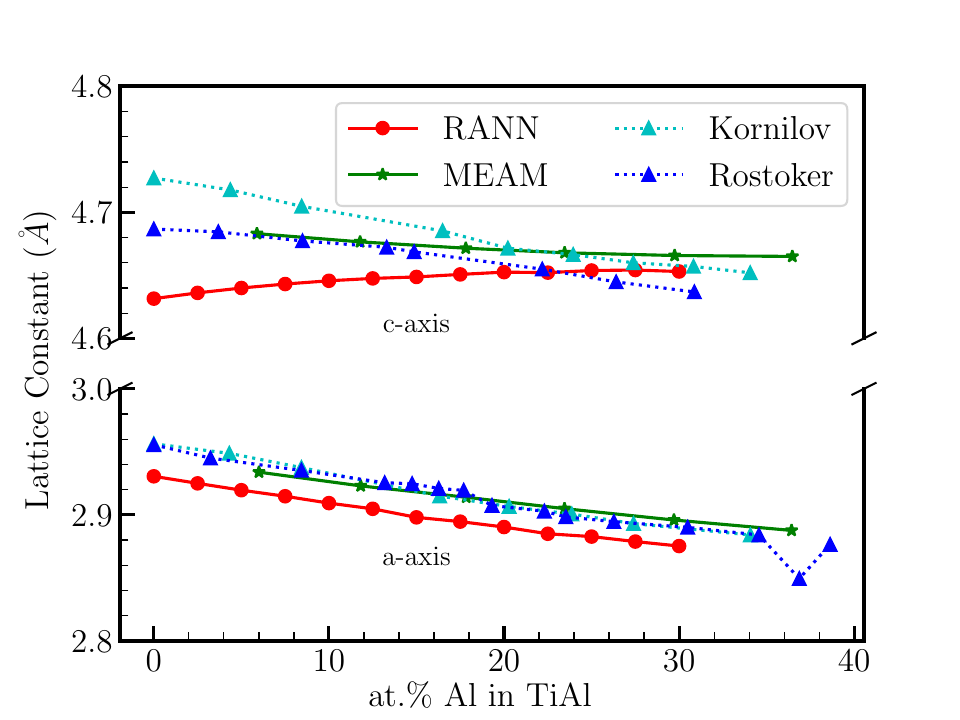}}
\caption{Lattic constant as a function of Al concentration for RANN (300 K), MEAM (0 K) \cite{kim2016atomistic}, and experimental values (300 K) \cite{kornilov1956phase,rostoker1952observations}.}
\label{lat const}
\end{figure}

Being able to accurately model the Ti-Al system with varying concentrations of Al is very important. A potential must accurately display fundamental material properties with varying concentrations of Al as well as predict at what temperature or Al concentration phase changes will occur. Both RANN and MEAM \cite{kim2016atomistic} are capable of predicting the lattice constants as a function of varying Al concentration in agreement with the experimental works of \citet{kornilov1956phase} and \citet{rostoker1952observations}. Lattice constants with varying aluminum temperature for MEAM at 0 K, RANN at 300 K, and experimental results at 300 K can be seen in \autoref{lat const}. MEAM shows more accuracy in this regard; however, RANN replicates the experimental trend for the a-axis and approaches the experimental results with increasing Al concentration for the c-axis.

\begin{figure}[!htbp]
\noindent \centering
\subfloat[900 K]{\includegraphics[width=1\columnwidth]{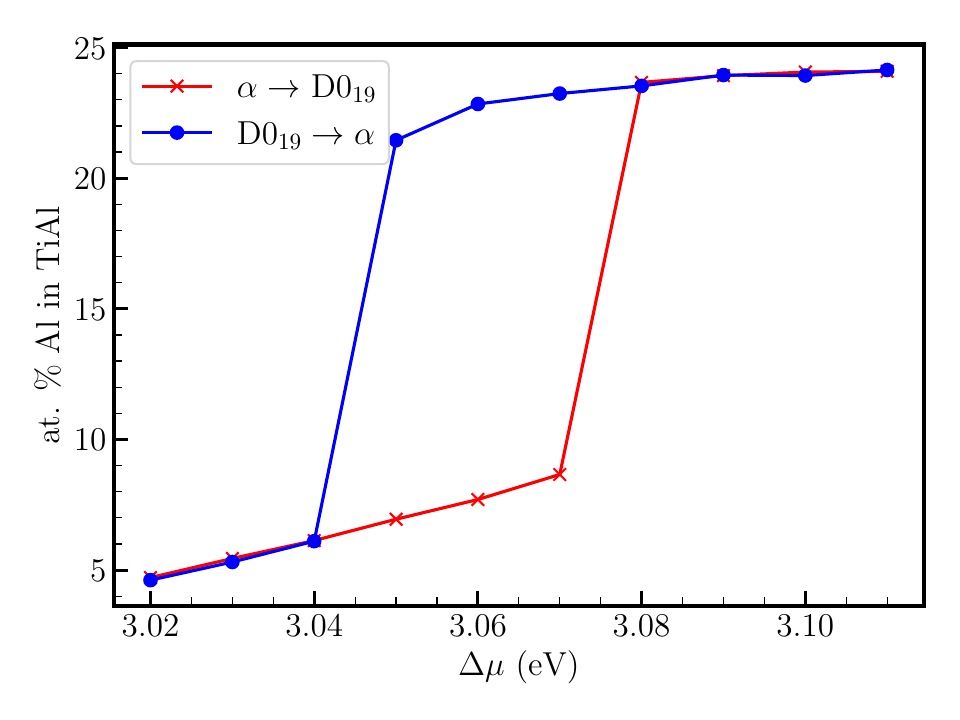}}
\caption{Hysteresis loop from Semi-Grand-Canonical Monte Carlo Simulations at 900 K.}
\label{mc}
\end{figure}

The RANN potential is capable of reproducing the phase diagram with up to 30$\%$ Al. The D0$_{19}$ to $\alpha$ transition predicted by RANN is comparable to that of CALPHAD as well as previous work done by \citet{kim2016atomistic}. Simulations for this transition were run at 900 K, 1000 K, 1100 K, and 1200 K. At temperatures higher than 1200 K the hysteresis loop becomes too small to accurately measure where the phase transition occurs. \autoref{mc} shows the hysteresis loop found at 900 K.

RANN predicts the $\alpha$ to $\alpha +$D0$_{19}$ phase boundary to occur at Al concentrations of 8.63\%, 9.66\%, 11.10\%, and 14.23\% for temperatures of 900 K, 1000 K, 1100 K, and 1200 K, respectively. RANN predicts the D0$_{19}$ to $\alpha +$D0$_{19}$ phase boundary to be at Al concentrations of 22.72\%, 22.25\%, 21.57\%, and 22.20\% for the same temperatures. While MEAM \cite{kim2016atomistic} displays an accurate transition between $\alpha$ and D0$_{19}$, MTP \cite{qi2023machine} does not accurately predict this phase transition. MTP predicts a stable phase between the $\alpha$ and D0$_{19}$ phases and can be seen in the supplementary information \cite{supp}.

\begin{figure}[!htbp]
\noindent \centering
\subfloat{\includegraphics[width=0.5\columnwidth]{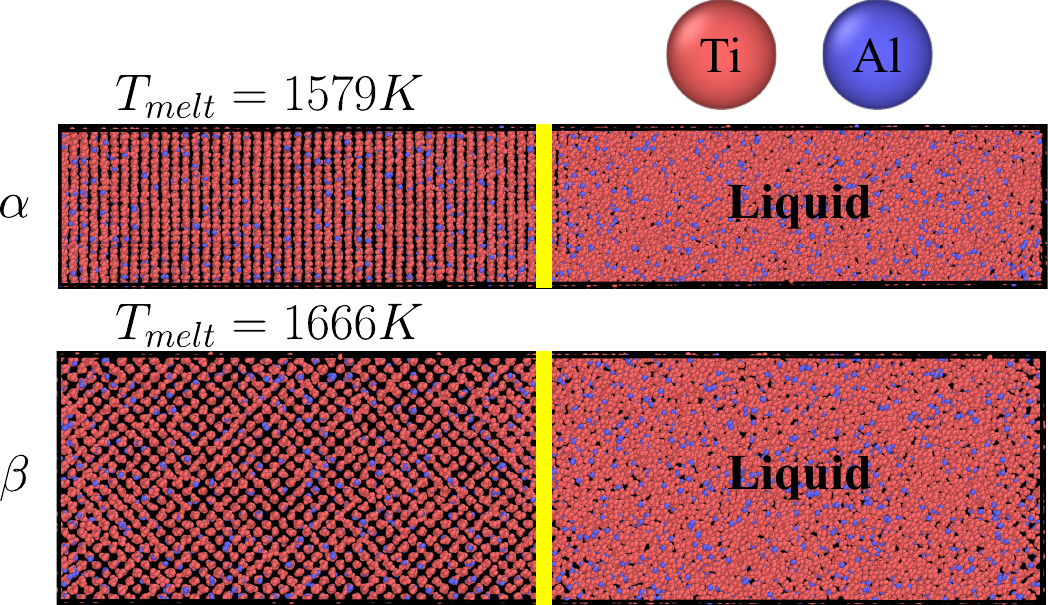}}
\subfloat{\includegraphics[width=0.5\columnwidth]{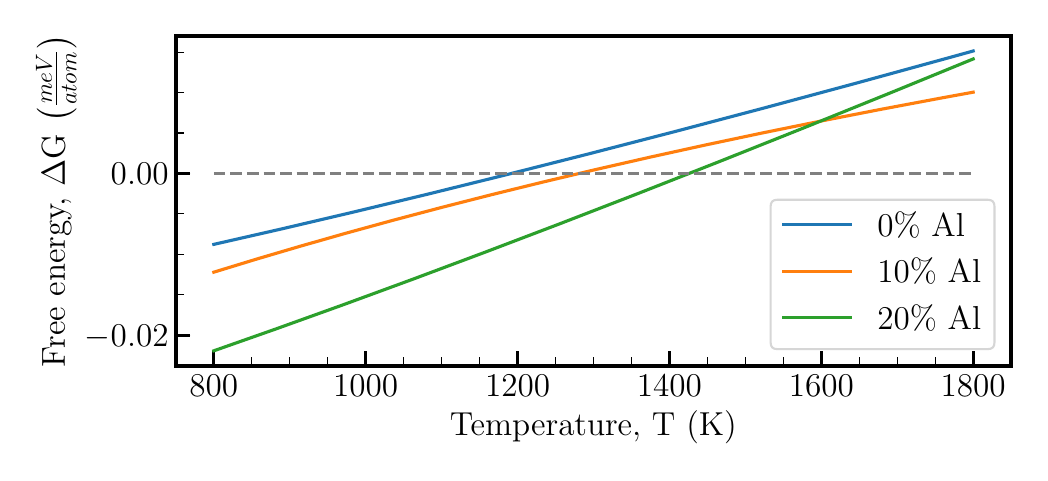}}
\caption{Left: solid-liquid interface for the $\alpha$ and $\beta$ phases for 10$\%$ Al concentration \cite{ovito}. Right: Gibbs free energy calculation for 0$\%$, 10$\%$, and 20$\%$ Al concentrations.}
\label{free energy}
\end{figure}

To the best of the authors' knowledge, no existing interatomic potential exists that accurately shows Al as an $\alpha$-stabilizer in Ti-Al alloys by reproducing the $\alpha$ to $\beta$ transition. The ability to model this phase transition is of great importance as many commercially available Ti-Al alloys contain two-phase regions composed of the $\alpha$ and $\beta$ phases in order to increase strength and creep resistance \cite{hennig2008classical}. Additionally, the $\beta$ phase is known to increase the workability of Ti alloys \cite{veiga2012properties}. \autoref{free energy} displays the solid-liquid interface for the $\alpha$ and $\beta$ phases at 10$\%$ Al concentration as well as the transition temperature for 0$\%$, 10$\%$, and 20$\%$ Al concentrations. Images used for the solid-liquid interface were taken from Ovito \cite{ovito}.

RANN predicts the $\alpha$ to $\beta$ transition to be along the upper boundary of the two-phase region predicted by CALPHAD. The transition temperature was calculated for Al concentrations ranging from 0\% Al to 20\% Al in 2.5\% increments.
\begin{table}[!htbp]
\caption{Melting temperatures used for calculation of the Gibbs relative free energy and $\alpha$-$\beta$ transition temperatures found from the Gibbs relative free energy.}
\label{melt transition temps}
\resizebox{\columnwidth}{!}{%
\begin{tabular*} {1\columnwidth}{@{\extracolsep{\fill}} lccc}

\hline \hline
at. $\%$ Al               & $T^{\alpha}_{melt}$ (K) & $T^{\beta}_{melt}$ (K) &  $T^{\alpha -\beta}_{transition}$ (K)\\
\hline
0	& 1525.295	& 1657.246	& 1192.05\\
2.5	& 1522.220	& 1672.751	& 1205.65\\
5	& 1534.717	& 1674.698	& 1235.20\\
7.5	& 1561.384	& 1691.315	& 1246.50\\
10	& 1579.092	& 1665.697	& 1280.20\\
12.5	& 1592.222	& 1700.295	& 1326.95\\
15	& 1618.833	& 1706.940	& 1362.04\\
17.5	& 1620.266	& 1717.814	& 1417.22\\
20	& 1633.606	& 1737.773	& 1426.22\\
\hline \hline
\end{tabular*}%
}
\end{table}

\begin{figure}[!htbp]
\noindent \centering
\subfloat{\includegraphics[width=1\columnwidth]{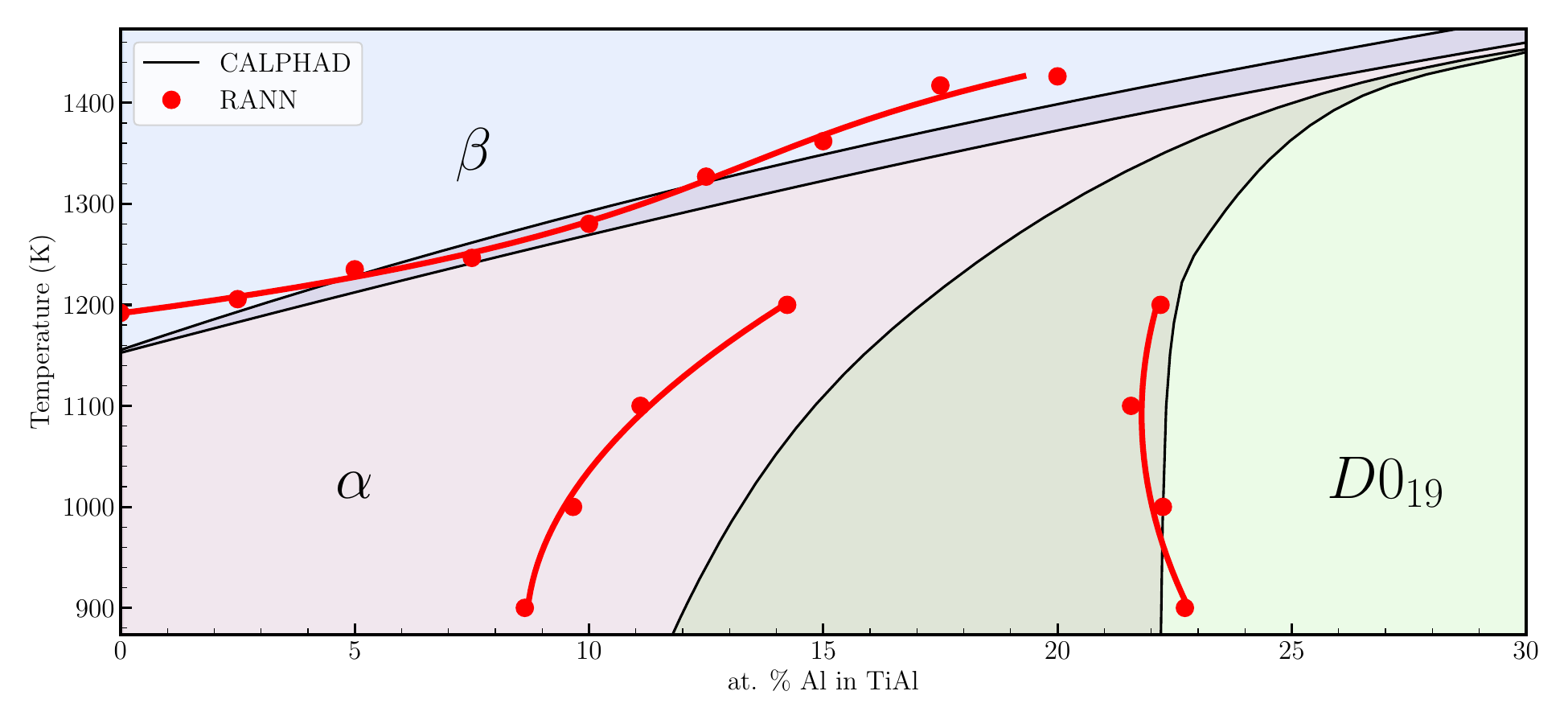}}
\caption{Phase diagram of Ti-Al binary system predicted by RANN compared to CALPHAD. The line between RANN data points is a cubic spline with an applied smoothing factor to indicate the trend of the data.}
\label{phase}
\end{figure}

In increasing order starting at 0\% Al, the transition temperatures are found to be 1192 K, 1206 K, 1235 K, 1247 K, 1280 K, 1327 K, 1362 K, 1417 K, and 1427 K. These values show a clear trend of increasing transition temperature with increasing Al concentration as is expected when adding an $\alpha$-stabilizer such as Al. The phase boundaries predicted by RANN are shown in \autoref{phase}.
\section{Discussion}
The microstructure of Ti-Al alloys plays a key role in determining material properties. Previous attempts at modeling the Ti-Al system have been limited to only capturing solid solutions or formed intermetallics -- not both. The RANN potential presented in this work bridges the gap between the currently available classical and ML potentials for the Ti-Al binary system. Classical potentials such as the MEAM potential introduced by \citet{kim2016atomistic} accurately model basic properties for solid solutions and formed intermetallic structures; however, they fail to capture accurate GSFEs and elastic constants for formed intermetallics. MEAM \cite{kim2016atomistic} also does not report on the $\alpha$-$\beta$ phase boundary. The MTP potential produced by \citet{qi2023machine} is promising for basic properties and plastic properties for formed intermetallic structures; however, it does not give accurate results for solid solutions. This ML potential also does not report on the $\alpha$-$\beta$ or $\alpha$-D0$_{19}$ phase boundaries. The Ti-Al RANN potential gives basic properties and GSFEs in agreement with DFT and phase boundaries in agreement with CALPHAD. The lattice constants and elastic constants predicted by RANN show an average deviation of 3.33$\%$ from DFT with the largest deviation coming from the $C_{44}$ elastic constant for $\alpha$-Ti. DFT shows a decrease in GSFE compared to $\alpha$-Ti when adding an Al atom adjacent to the glide plane. RANN accurately displays this trend as well. RANN mimics the slope of the phase boundaries between $\alpha$ and $\beta$ as well as $\alpha$ and D0$_{19}$ given by CALPHAD. The $\alpha$ to D0$_{19}$ phase boundary predicted by RANN occurs at an Al concentration roughly 3$\%$ lower than that of CALPHAD; however the D0$_{19}$ to $\alpha$ phase boundary given from RANN is within approximately 1$\%$ of CALPHAD. As the temperature increases in systems with an Al concentration near the $\alpha$ to D0$_{19}$ phase boundary, RANN correctly predicts that the $\alpha$ phase becomes more stable than the two-phase region containing $\alpha$ and D0$_{19}$. RANN shows the $\alpha$-$\beta$ boundary to be along the upper phase boundary given by CALPHAD. This shows that the RANN potential accurately finds Al to be an $\alpha$ stabilizer when added to Ti. These results show that RANN is capable of modeling phase behaviors, solute effects, and thermomechanical properties of Ti-Al systems with up to 30$\%$ Al concentration. Furthermore, to the authors' knowledge, this is the only interatomic potential that is capable of reproducing the $\alpha$-$\beta$ phase boundary in the Ti-Al binary system. 
\clearpage
\newpage
\bibliography{TiAl_ref}
\end{document}


\title{Predicting Ti-Al Binary Phase Diagram with an Artificial Neural Network Potential - Supplemental Information}
\author{Micah Nichols}
\affiliation{Michael W. Hall School of Mechanical Engineering}
\affiliation{Center for Advanced Vehicular Systems, Mississippi State University, Mississippi State, Mississippi 39762, USA}
\author{Mashroor S. Nitol}
\affiliation{Los Alamos National Laboratory, Los Alamos, NM, 87545, USA}
\author{Saryu J. Fensin}
\affiliation{Los Alamos National Laboratory, Los Alamos, NM, 87545, USA}
\author{Christopher D. Barrett}
\affiliation{Michael W. Hall School of Mechanical Engineering}
\affiliation{Center for Advanced Vehicular Systems, Mississippi State University, Mississippi State, Mississippi 39762, USA}
\author{Doyl E. Dickel}
\email{doyl@me.msstate.edu}
\affiliation{Michael W. Hall School of Mechanical Engineering}
\affiliation{Center for Advanced Vehicular Systems, Mississippi State University, Mississippi State, Mississippi 39762, USA}

\maketitle
\newpage
\section{DFT training database}
The DFT training database contains 69,558 simulations with 3,140,478 unique atomic environments. Simulations can be broken down into those that use bulk structures and those that use structures with defects. Both of these groups contain structures of pure Ti, pure Al, and Ti-Al. Bulk structures are pristine crystalline structures that are strained and undergo thermal perturbation. Simulations for bulk structures can be found in \autoref{table:database1}. Defect structures contain interstitials, vacancies, dislocation cores, or other imperfections within the crystal lattice. For the purpose of this database, amorphous data is also considered a defect structure. Information on defect structures in the DFT database can be found in \autoref{table:database2}.
\begingroup \squeezetable
\begin{table*}[!htbp]
\caption{\label{table:database1} DFT database of bulk structures used for features in input layers of the RANN potential.}
\centering
 \begin{adjustbox}{max width=\columnwidth,center}
\begin{ruledtabular}
\begin{tabular}{l c c c c} 
Sample description & \multicolumn{1} {p{2cm}} {\centering Atoms \\ per \\ simulation} & \multicolumn{1} {p{2cm}} {\centering Number \\of \\ simulations}  & \multicolumn{1} {p{2cm}} {\centering Total \\ atomic \\ environment}  \\ 
\hline
 $\alpha-$Ti w$/$ volumetric strains up to $\pm$ 20\% &  2  &  99 & 198 \\
  $\beta$-Ti w$/$ volumetric strains up to $\pm$ 20\% &  2  &  99 & 198 \\
  $\omega$-Ti w$/$ volumetric strains up to $\pm$ 20\% &  3  &  99 & 297 \\
 $a15-$Ti w$/$ volumetric strains up to $\pm$ 20\% &  8  &  99 & 792 \\
  $DC$-Ti w$/$ volumetric strains up to $\pm$ 20\% &  8  &  99 & 792 \\
  $FCC$-Ti w$/$ volumetric strains up to $\pm$ 20\% &  4  &  99 & 396 \\ 
$sc$-Ti w$/$ volumetric strains up to $\pm$ 20\% &  1  &  99 & 99 \\ 
   $FCC$-Al w$/$ volumetric strains up to $\pm$ 20\% &  2  &  99 & 198 \\
  $\omega$-Al w$/$ volumetric strains up to $\pm$ 20\% &  3  &  99 & 297 \\ 
  
  $\alpha-$Ti w$/$ shear strains up to $\pm$ 5\% &  2  &  1386 & 2772 \\
    $\beta-$Ti w$/$ shear strains up to $\pm$ 5\% &  2  &  1386 & 2772 \\
      $\omega-$Ti w$/$ shear strains up to $\pm$ 5\% &  3  &  1386 & 4158 \\
    $FCC-$Al w$/$ shear strains up to $\pm$ 5\% &  2  &  1386 & 2772 \\
$Ti_{3}Al$ (DO$_{19}$) w$/$ shear strains up to $\pm$ 5\% &  8  &  1386 & 10848 \\
$TiAl$ (L1$_{0}$) w$/$ shear strains up to $\pm$ 5\% &  4  &  1256 & 5024 \\
    $FCC-$Al w$/$ shear strains up to $\pm$ 5\% &  2  &  1386 & 2772 \\  
$\alpha-Ti$ $3\times3\times3$ orthogonal supercell w$/$ strains up to $\pm$ 5\% &  54  &  961 & 51894 \\     
$\beta-Ti$ $3\times3\times3$ orthogonal supercell w$/$ strains up to $\pm$ 5\% &  54  &  1000 & 54000 \\    
$\omega-Ti$ $3\times3\times3$ orthogonal supercell w$/$ strains up to $\pm$ 5\% &  81  &  783 & 63423 \\ 
$FCC-Ti$ $3\times3\times3$ orthogonal supercell w$/$ strains up to $\pm$ 5\% &  32  &  300 & 9600 \\ 
$FCC-Al$ $3\times3\times3$ orthogonal supercell w$/$ strains up to $\pm$ 5\% &  54  &  995 & 53730 \\
$FCC-Al$ $3\times3\times3$ orthogonal supercell w$/$ strains up to $\pm$ 5\% &  32  &  299 & 9568 \\
$HCP-Al$ $3\times3\times3$ orthogonal supercell w$/$ strains up to $\pm$ 5\% &  54  &  135 & 7290 \\
$\omega-Al$ $2\times2\times3$ orthogonal supercell w$/$ strains up to $\pm$ 5\% &  36  &  199 & 7164 \\
   $Ti_{3}Al$ (DO$_{19}$) $2\times2\times2$ orthogonal supercell   w$/$ strains up to $\pm$ 5\% &  64  &  500 & 38336 \\
$TiAl(L1_{0})$ $2\times2\times2$ orthogonal supercell   w$/$ strains up to $\pm$ 5\% &  32  &  500 & 16000 \\
   $TiAl_{3}$ (DO$_{22}$ $2\times2\times2$ orthogonal supercell   w$/$ strains up to $\pm$ 5\% &  64  &  500 & 38336 \\
   $TiAl_{2}$ (Ga$_{2}$Hf) $2\times2\times1$ orthogonal supercell   w$/$ strains up to $\pm$ 5\% &  96  &  500 & 48000 \\   
$TiAl(B2)$ $2\times2\times2$ orthogonal supercell   w$/$ strains up to $\pm$ 5\% &  108  &  293 & 31644 \\
$\beta-Ti+Al(1-50\%)$ $3\times3\times3$ orthogonal supercell w$/$ strains up to $\pm$ 5\% &  54  &  1636 & 88344 \\
$\alpha-Ti+Al(1-50\%)$ $3\times3\times3$ orthogonal supercell w$/$ strains up to $\pm$ 5\% &  54  &  1899 & 102546 \\
$\omega-Ti+Al(1-50\%)$ $3\times3\times2$ orthogonal supercell w$/$ strains up to $\pm$ 5\% &  54  &  966 & 53784 \\
$FCC-Ti+Al(1-50\%)$ $2\times2\times2$ orthogonal supercell\\ w$/$ strains up to $\pm$ 5\% &  32  &  762 & 24384 \\
$FCC-Al+Ti(1-50\%)$ $2\times2\times2$ orthogonal supercell\\ w$/$ strains up to $\pm$ 5\% &  32  &  1799 & 57568 \\
$HCP-Al+Ti(1-50\%)$ $3\times3\times3$ orthogonal supercell\\ w$/$ strains up to $\pm$ 5\% &  54  &  852 & 46008 \\

 \end{tabular}
 \end{ruledtabular}
    \end{adjustbox}
\end{table*}
\begin{table*}[!htbp]
\caption{\label{table:database2} DFT database of defects used for features in input layers of the RANN potential.}
\centering
 \begin{adjustbox}{max width=.8\columnwidth,center}
\begin{ruledtabular}
\begin{tabular}{l c c c c} 
Sample description & \multicolumn{1} {p{2cm}} {\centering Atoms \\ per \\ simulation} & \multicolumn{1} {p{2cm}} {\centering Number \\of \\ simulations}  & \multicolumn{1} {p{2cm}} {\centering Total \\ atomic \\ environment}  \\ 
\hline
$\alpha-Ti$(1 Octahedral interstitial) $3\times3\times3$ orthogonal supercell\\ w$/$ strains up to $\pm$ 5\% &  55  &  89 & 4895 \\
$\beta-Ti$(1 Octahedral interstitial) $3\times3\times3$ orthogonal supercell\\ w$/$ strains up to $\pm$ 5\% &  55  &  201 & 11055 \\
$FCC-Al$(1 Octahedral interstitial) $3\times3\times3$ orthogonal supercell\\ w$/$ strains up to $\pm$ 5\% &  55  &  201 & 11055 \\
$\alpha-Ti$(1 Octahedral interstitial)$+Al$:$1-50\%)$ $3\times3\times3$ orthogonal supercell\\ w$/$ strains up to $\pm$ 5\% &  55  &  420 & 23100 \\
$\beta-Ti$(1 Octahedral interstitial)$+Al$:$1-50\%)$ $3\times3\times3$ orthogonal supercell\\ w$/$ strains up to $\pm$ 5\% &  55  &  1200 & 66000 \\
$FCC-Al$(1 Octahedral interstitial)$+Ti$:$1-50\%)$ $2\times2\times2$ orthogonal supercell\\ w$/$ strains up to $\pm$ 5\% &  33  &  1129 & 37257 \\
$\alpha-Ti$(1 Tetrahedral interstitial) $3\times3\times3$ orthogonal supercell\\ w$/$ strains up to $\pm$ 5\% &  55  &  199 & 10945 \\
$\beta-Ti$(1 Tetrahedral interstitial) $3\times3\times3$ orthogonal supercell\\ w$/$ strains up to $\pm$ 5\% &  55  &  201 & 11055 \\
$FCC-Al$(1 Tetrahedral interstitial) $3\times3\times3$ orthogonal supercell\\ w$/$ strains up to $\pm$ 5\% &  33  &  201 & 6633 \\
$\beta-Ti$(1 Tetrahedral interstitial)$+Al$:$1-50\%)$ $3\times3\times3$ orthogonal supercell\\ w$/$ strains up to $\pm$ 5\% &  55  &  1198 & 65890 \\
$FCC-Al$(1 Tetrahedral interstitial)$+Ti$:$1-50\%)$ $3\times3\times3$ orthogonal supercell\\ w$/$ strains up to $\pm$ 5\% &  33  &  1200 & 39600 \\
$\alpha-Ti$(Mono vacancy) $3\times3\times3$ orthogonal supercell w$/$ strains up to $\pm$ 5\% &  53  &  117 & 6201 \\
$\beta-Ti$(Mono vacancy) $3\times3\times3$ orthogonal supercell w$/$ strains up to $\pm$ 5\% &  53  &  200 & 10600 \\
$\omega-Ti$(Mono vacancy) $2\times2\times3$ orthogonal supercell w$/$ strains up to $\pm$ 5\% &  53  &  106 & 5618 \\
$FCC-Al$(Mono vacancy) $3\times3\times3$ orthogonal supercell\\ w$/$ strains up to $\pm$ 5\% &  31  &  200 & 6200 \\
$\alpha-Ti$(Mono vacancy)$+Al$:$1-50\%)$ $3\times3\times3$ orthogonal supercell\\ w$/$ strains up to $\pm$ 5\% &  53  &  550 & 29150 \\
$\beta-Ti$(Mono vacancy)$+Al$:$1-50\%)$ $3\times3\times3$ orthogonal supercell\\ w$/$ strains up to $\pm$ 5\% &  53  &  1200 & 63600 \\
$FCC-al$(Mono vacancy)$+Ti$:$1-50\%)$ $3\times3\times3$ orthogonal supercell\\ w$/$ strains up to $\pm$ 5\% &  33  &  1180 & 38940 \\
$\alpha-Ti$(\hkl[0001],\hkl[1100], \hkl[1210]) monolayer  orthogonal supercell\\ w$/$ strains up to $\pm$ 5\% &  (32-36)  &  580 & 16120 \\
$\beta-Ti$(\hkl[100],\hkl[110],\hkl[112]) monolayer  orthogonal supercell\\ w$/$ strains up to $\pm$ 5\% &  (32-48)  &  590 & 23600 \\
$\omega-Ti$(\hkl[0001]) monolayer  orthogonal supercell w$/$ strains up to $\pm$ 5\% &  36  &  200 & 7200 \\
$FCC-Al$(\hkl[100],\hkl[110],\hkl[112]) monolayer  orthogonal supercell\\ w$/$ strains up to $\pm$ 5\% &  (32-48)  &  346 & 11764 \\
$\alpha-Ti$(\hkl[0001],\hkl[1100], \hkl[1210])$+Al$:$1-25\%)$ monolayer  orthogonal supercell\\ w$/$ strains up to $\pm$ 5\% &  (32-36)  &  983 & 33422 \\
$\beta-Ti$(\hkl[100],\hkl[110],\hkl[112])$+Al$:$1-25\%)$ monolayer  orthogonal supercell\\ w$/$ strains up to $\pm$ 5\% &  (32-48)  &  882 & 35280 \\
$\omega-Ti$(\hkl[0001])$+Al$:$1-25\%)$ monolayer  orthogonal supercell\\ w$/$ strains up to $\pm$ 5\% &  36  &  352 & 12672 \\
$FCC-Al$(\hkl[100],\hkl[110],\hkl[112])$+Ti$:$1-25\%)$ monolayer  orthogonal supercell\\ w$/$ strains up to $\pm$ 5\% &  (32-48)  &  724 & 24616 \\
$\alpha-Ti$(\hkl[0001],\hkl[1100],\hkl[1120]) separation distance  &  (16-48)  &  200 & 6400 \\
$FCC-Al$(\hkl[100],\hkl[110],\hkl[123]) separation distance  &  (12-36)  &  200 & 4800 \\
$\alpha-Ti$ dislocation core structure\\ (prismatic,prismatic-saddle,pyramidal, pyramidal-saddle)  &  192  &  80 & 15360 \\
Amorphous structure : $Ti$& (40-43) & 200 & 8300\\
Amorphous structure : $Al$& (55-62) & 200 & 11700\\
Amorphous structure : $Ti-Al(1-50\%) $& (66-95) & 200 & 57892\\
Isolate atom & 2 & 1 & 2 \\
 \end{tabular}
 \end{ruledtabular}
    \end{adjustbox}
\end{table*}
\endgroup
\clearpage
\section{Ti}
The values for the material properties of the $\alpha$, $\beta$, and $\omega$ phases of pure Ti from experiments, DFT, and RANN at 0 K can be seen in \autoref{pure ti table}. This table shows that the Ti-Al RANN potential gives results near the accuracy of DFT.
\begin{table}[!htbp]
\caption{Pure Ti properties: \emph{italic} are from previous DFT. $a$ and $c$ - lattice constants, $E_c$ - cohesive energy, $B$ - the bulk modulus, $C$ - elastic constants, $\Delta E$ - structural energy difference, $E_{surface}$ - surface energy, $E^{vac}$ - vacancy formation energy, $E^{inter}$ - interstitial energy. Experimental results for the $\beta$ phase are at finite temperature.}
\label{pure ti table}
\resizebox{\columnwidth}{!}{%
\begin{tabular*} {1.2\textwidth}{@{\extracolsep{\fill}} ccccccc} 
\hline \hline
& \multicolumn{2}{c}{$\alpha$}                  & \multicolumn{2}{c}{$\beta$} & \multicolumn{2}{c}{$\omega$} \\ \hline
         Properties            & Expt./DFT & RANN   & Expt./DFT      & RANN        & DFT       & RANN         \\ \hline
         $a$ ($\AA$)                     & 2.947$^a$/\emph{2.92$^b$}, 2.92    & 2.93   &   \emph{3.26$^c$},3.24        & 3.24        &  4.55          & 4.53        \\
         $c$ ($\AA$)                     & 4.686$^a$/\emph{4.63$^b$}, 4.63    & 4.61   &           &             &      2.82      & 2.83         \\
         $c/a$ ratio             & 1.586$^a$/\emph{1.58$^b$}, 1.58    & 1.58   &           &             &     0.62       & 0.62         \\
         $E_{c}$ ($eV$)                    & 4.85$^d$/5.32    & 3.23   &   5.22        & 3.13        &    5.31        & 3.24         \\
         $B$ ($GPa$)                     &         & 117.59 &           & 110.93      &            & 115.73       \\
         $C_{11}$ ($GPa$)                   & 176.1$^a$/\emph{177$^b$},176.6   & 170.41 &   134$^e$/\emph{89$^c$},93.51      & 124.36       &    \emph{198$^c$},193.15       & 178.29       \\
         $C_{12}$ ($GPa$)                   & 86.9$^a$/\emph{90$^b$},91.6      & 91.45  &   110$^e$/\emph{112$^c$},114.6     &  108.63      &    \emph{84$^c$},86        & 112.87       \\
         $C_{13}$ ($GPa$)                   & 68.3$^a$/\emph{84$^b$}, 84.2    & 83.56 & & &  \emph{53$^c$},54 & 79.65\\
         $C_{33}$ ($GPa$)                   & 190.5$^a$/\emph{189$^b$}, 192.2     & 197.19 &           &             &   \emph{246$^c$},248        & 206.35      \\
         $C_{44}$ ($GPa$)                   & 50.8$^a$/\emph{40$^b$}, 40.72    & 47.75  &     36$^e$/\emph{42$^c$},40.19      & 44.53       &    \emph{54$^c$},54        & 65.97       \\
         $\Delta E_{hcp\rightarrow bcc}$ ($\frac{eV}{atom}$)     & /0.10   & 0.10  &           &             &            &              \\
         $\Delta E_{hcp\rightarrow fcc}$ ($\frac{eV}{atom}$)     & /0.06   & 0.06  &           &             &            &              \\
         $\Delta E_{hcp\rightarrow sc}$ ($\frac{eV}{atom}$)      & /0.79  & 0.79  &           &             &            &              \\
         $\Delta E_{hcp\rightarrow a15}$ ($\frac{eV}{atom}$)     & /0.18  & 0.18  &           &             &            &              \\
         $\Delta E_{hcp\rightarrow dc}$ ($\frac{eV}{atom}$)      & /2.266   & 1.95  &           &             &            &              \\
         $\Delta E_{hcp\rightarrow \omega}$ ($\frac{eV}{atom}$)   & /-0.01  & -0.01 &           &             &            &              \\
         $E_{surface}^{\hkl(0001)}$ ($\frac{mJ}{m^2}$)          &     2.1$^f$/1.95    &  1.93      &           &             &            &              \\
         $E_{surface}^{\hkl(1010)}$ ($\frac{mJ}{m^2}$)         &   1.92$^g$/2.00      &    2.01    &           &             &            &              \\
         $E_{f}^{vac}$ ($eV$)               &   /2.03        &  2.04    &     &             &            &              \\
         $E_{octa}^{inter}$ ($eV$)            &    /2.59     &   2.62     &           &             &            &       \\ 
         $E_{Dumbbell\left[0001\right]}^{inter}$ ($eV$)            &    /2.87     &  2.85      &           &             &            &       \\        \hline \hline  
         \multicolumn{7}{l}{\small $^{a}$\cite{simmons1971single}, $^{b}$\cite{yin2017comprehensive}, $^{c}$\cite{nitol2022machine}, $^{d}$\cite{kittel2005introduction}, $^{e}$\cite{fisher1964single}, $^{f}$\cite{boer1988cohesion}, $^g$\cite{tyson1977surface} } 
\end{tabular*}%
}
\label{ti_table}
\end{table}
\section{Al}
\begin{figure}[!htbp]
\noindent \centering
\subfloat{\includegraphics[width=1\textwidth]{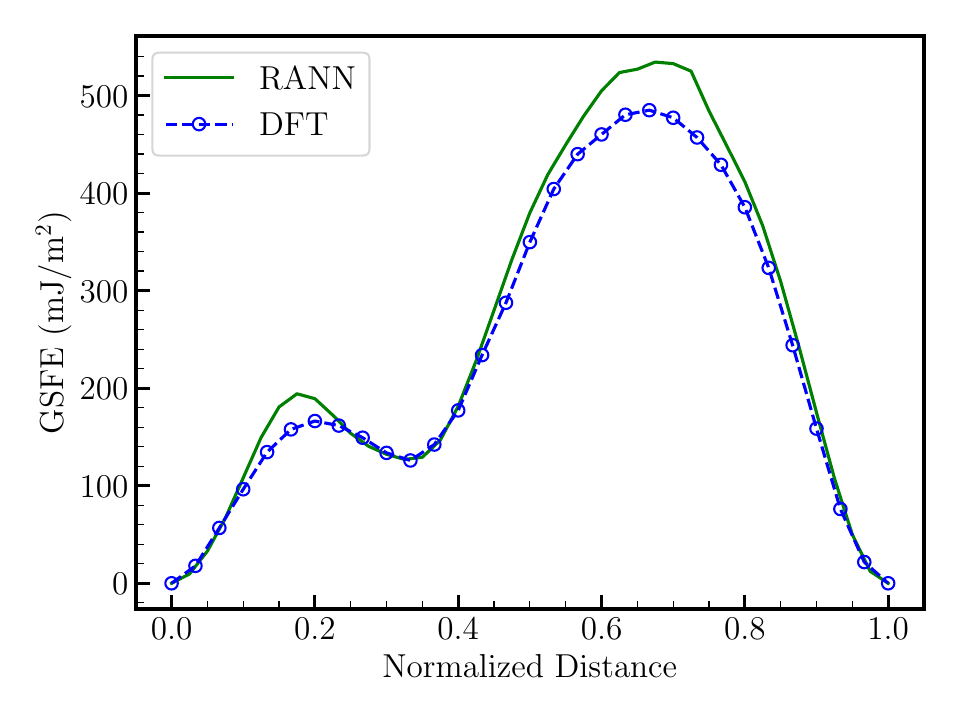}}\\
\caption{GSFE for pure Al along the \hkl[112] direction in the \hkl(111) plane}
\label{al_props}
\end{figure}

RANN predicts the material properties for pure Al to be near that of DFT calculations as seen in \autoref{pure al table}. RANN predictions for the generalized stacking fault energy (GSFE) along the \hkl[112] direction in the \hkl(111) plane can be seen in \autoref{al_props}.
\begin{table}[!htbp]
\caption{Pure Al properties: \emph{italic} are from previous DFT. $a$ - lattice constant, $E_c$ - cohesive energy, $B$ is the bulk modulus, $C$ - elastic constants, $\Delta E$ - structural energy difference, $\gamma_{usf}$ - unstable stacking fault energy, $\gamma_{ssf}$ stable stacking fault energy, $E_{surface}$ - surface energy, $E^{vac}$ - vacancy formation energy, $E^{inter}$ - interstitial energy.}
\label{pure al table}
\resizebox{\columnwidth}{!}{%
\begin{tabular*} {1.2\textwidth}{@{\extracolsep{\fill}} cccc}

\hline \hline
Properties               & Expt.         & DFT    & RANN   \\
\hline
$a$ ($\AA$)                        & 4.05$^a$        & 4.04   & 4.04      \\
        $E_{c}$ ($eV$)                       & 3.39$^b$        &    & 3.36       \\
        $B$ ($GPa$)                        & 72.2$^b$         & 77.11  &  75.01        \\
        $C_{11}$ ($GPa$)                      & 114.3$^b$       & \emph{119.2$^c$,114.6$^c$,122.2$^d$,110.4$^e$,109.3$^f$,101.0$^g$},110.93 & 94.63       \\
        $C_{12}$ ($GPa$)                      & 61.9$^b$         & \emph{64.3$^c$,63.2$^c$,60.8$^d$,54.5$^e$,57.5$^f$,61.0$^g$},58.2  & 65.20       \\
        $C_{44}$ ($GPa$)                      & 31.6$^b$         & \emph{36.5$^c$,35.3$^c$,37.4$^d$,31.3$^e$,30.1$^f$,25.4$^g$},32.2  & 46.14        \\
        $\Delta E_{fcc\rightarrow bcc}$ ($\frac{eV}{atom}$)        &              & 0.09   & 0.10          \\
        $\Delta E_{fcc\rightarrow hcp}$ ($\frac{eV}{atom}$)        &              & 0.03   & 0.03          \\
        $\Delta E_{fcc\rightarrow sc}$ ($\frac{eV}{atom}$)         &              & 0.37   & 0.37          \\
        $\Delta E_{fcc\rightarrow a15}$ ($\frac{eV}{atom}$)        &              & 0.08   & 0.08          \\
        $\Delta E_{fcc\rightarrow dc}$ ($\frac{eV}{atom}$)         &              & 0.74   & 0.75          \\
        $\Delta E_{fcc\rightarrow \omega}$ ($\frac{eV}{atom}$) &  & 0.11   & 0.11          \\
        $\gamma_{usf}^{\hkl(110)\hkl[112]}$ ($\frac{mJ}{m^2}$)  &             & \emph{224$^{h,i}$},158.2 & 171.34       \\
        $\gamma_{ssf}^{\hkl(110)\hkl[112]}$ ($\frac{mJ}{m^2}$) &    135-166$^{b,j}$            & \emph{122-164$^{h,i,k}$},125.9 & 123.88        \\
        $E_{surf}^{\hkl(001)}$ ($eV$)              &              & 873.8 & 926.69        \\
        $E_{surf}^{\hkl(110)}$ ($eV$)              &              & 927.7 & 968.80        \\
        $E_{surf}^{\hkl(111)}$ ($eV$)             &              & 820.2 & 846.27        \\
        $E_{f}^{vac}$ ($eV$)                  & 0.67$^l$         & 0.65   & 0.62          \\
        $E_{tetra}^{inter}$ ($eV$)              &              & 3.21   & 3.21          \\
        $E_{octa}^{inter}$ ($eV$)               &              & 2.84   & 2.92    \\
        \hline \hline
        \multicolumn{4}{l}{\small $^{a}$\cite{blakemore1985solid}, $^{b}$\cite{kittel2005introduction}, $^{c}$\cite{malica2020quasi}, $^{d}$\cite{li1998ab}, $^{e}$\cite{wang2009ab}, $^{f}$\cite{golesorkhtabar2013elastic}, $^g$\cite{shang2010first}, $^h$\cite{lu2000generalized}, $^i$\cite{wu2010ab}, $^j$\cite{smallman1970stacking}, $^k$\cite{woodward2008prediction,crampin1990calculation,denteneer1991defect,hammer1992stacking}, $^l$\cite{hehenkamp1994absolute} }
\end{tabular*}%
}
\label{al_table}
\end{table}
\clearpage
\section{Ti-Al}
\begin{figure}[!h]
\noindent \centering
\subfloat[$\alpha$ - Ti\hkl(0001)]{\includegraphics[width=0.33\textwidth]{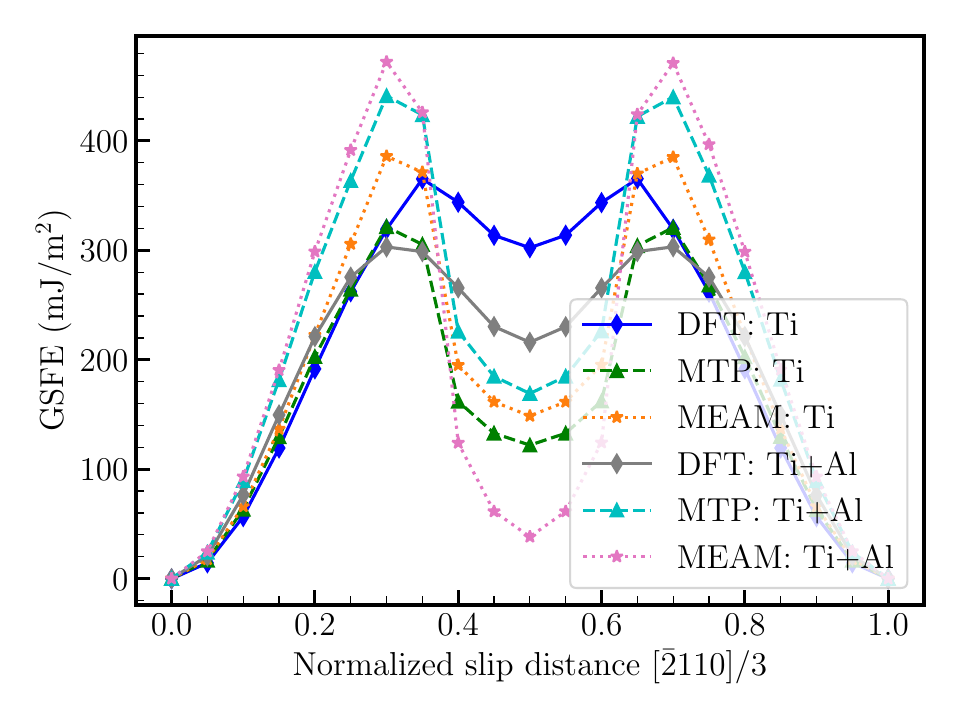}}
\subfloat[$\alpha$ - Ti\hkl(0001)]{\includegraphics[width=0.33\textwidth]{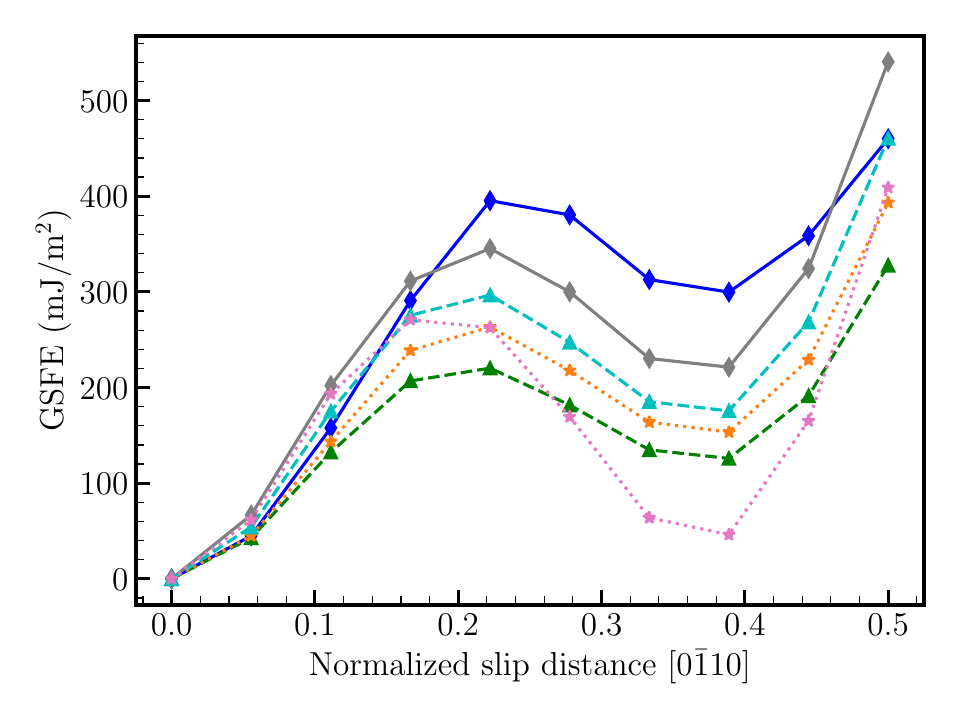}}
\subfloat[$\alpha$ - Ti\hkl(1010)]{\includegraphics[width=0.33\textwidth]{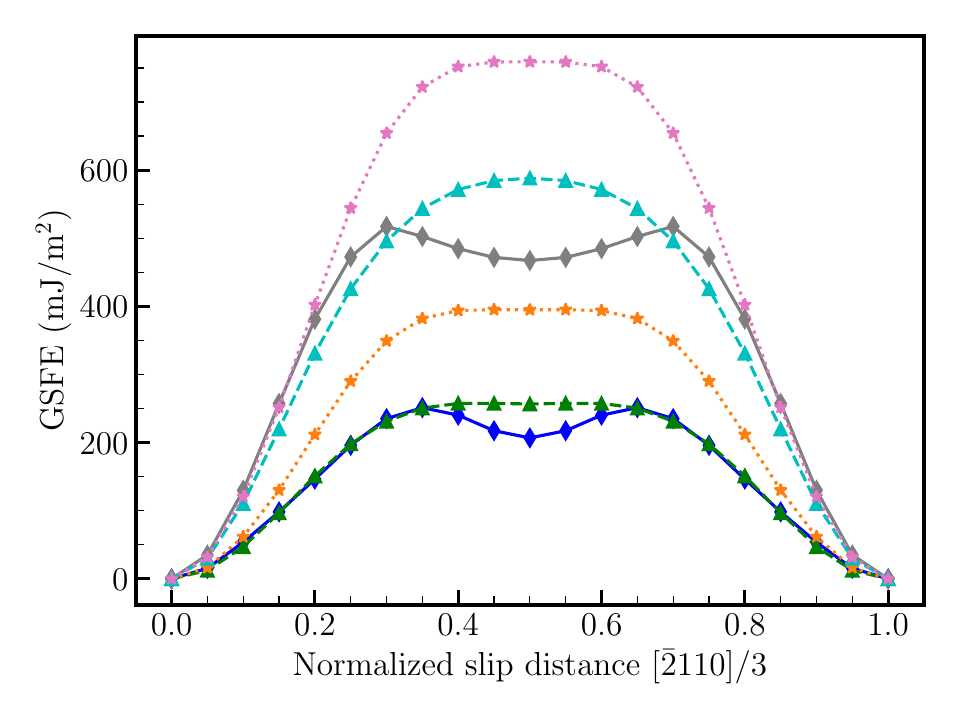}}
\caption{GSFE of Ti-Al alloy for DFT and MTP, where solute Al's position is in the first nearest positions relative to the glide plane)}
\label{mtp meam gsfe}
\end{figure}

While MEAM \cite{kim2016atomistic} captures basic solid solution properties well, it fails to accurately predict plastic properties, such as the GSFE, for solid solutions. MTP \cite{qi2023machine} does well with GSFEs for formed intermetallics, but it also struggles with finding correct GSFEs for solid solutions. When adding an Al atom adjacent to the glide plane in the $\alpha$ phase, the GSFE is expected to decrease compared to that of pure $\alpha$-Ti. The GSFE of pure $\alpha$-Ti and Ti with one Al atom adjacent to the glide plane were calculated for both MEAM \cite{kim2016atomistic} and MTP. \cite{qi2023machine}. Structures ranging from 20-80 atoms with a fault plane approximately in the middle of the cell were created with the atomman Python package \cite{BECKER2013277,Hale_2018}. Atoms above the fault plane were incrementally shifted in the \emph{x} direction until the structure was back in the original configuration. Each structure was placed in a LAMMPS simulation with periodic boundary conditions in the \emph{x} and \emph{y} directions and a free surface in the \emph{z} direction. The simulation fixed atoms in the \emph{x} and \emph{y} directions but allowed relaxation in the direction normal to the glide plane (\emph{z} direction). Once the GSFE for pure $\alpha$-Ti was calculated, one Ti atom was swapped to be an Al atom. \autoref{mtp meam gsfe} shows GSFE results for MEAM \cite{kim2016atomistic} and MTP \cite{qi2023machine} compared to DFT. MEAM shows this decrease in the GSFE, but underestimates the GSFE in the basal and prismatic planes. MTP predicts a higher GSFE when an Al atom is introduced adjacent to the glide plane. RANN shows both more accurate GSFEs and a decrease in the GSFE with the addition of an Al atom. This can be seen in Fig. 5 of the main article.

The convex hull for RANN seen in Fig. 3 of the main text was found by allowing Ti-Al structures of varying space groups to relax in a LAMMPS simulation. \autoref{convex_table} displays energies from DFT and RANN for all structures and space groups tested.
\begin{table}[!htbp]
\caption{Formation energies of selected ordered structures  (units of meV/atom) from both DFT and RANN potential.}
\resizebox{\columnwidth}{!}{%
\label{convex_table}
\begin{tabular*} {1.4\textwidth}{@{\extracolsep{\fill}} llcc||llcc}
\hline \hline
Structure              & Space group      & DFT  & RANN & Structure              & Space group      & DFT  & RANN \\ \hline
\ch{Ti5Al}             & $P6_{3}2\bar{2}$      & -137 & -126 & TiAl($B2$)             & $Pm\bar{3}m$      & -262 & -290 \\
\ch{Ti4Al}             & $I4/m$                & -179 & -165 & TiAl ($B8_{1}$)        & $P6_{3}/mmc$      & -252 & -261 \\
\ch{Ti3Al}($D0_{19}$)  & $P6_{3}/mmc$          & -281 & -283 & TiAl($B_{h}$)          & $P\bar{6}m2$      & -251 & -244 \\
\ch{Ti3Al}($D0_{24}$)  & $P6_{3}/mmc$          & -268 & -272 & TiAl                   & $Fm\bar{3}m$      & -173 & -207 \\
\ch{Ti3Al}($L1_{2}$)   & $Pm\bar{3}m$          & -264 & -262 & TiAl($B20$)            & $P2_{1}3$         & -120 & -134 \\
\ch{Ti3Al}($D0_{22}$)  & $I4/mmm$              & -254 & -243 & TiAl ($B32$)           & $Fd\bar{3}m$      & -72  & -95  \\
\ch{Ti3Al}($D0_{23}$)  & $I4/mmm$              & -259 & -251 & \ch{Ti2Al3}            & $P4/m$            & -419 & -418 \\
\ch{Ti3Al} ($L2_{1})$  & $Fm\bar{3}m$          & -144 & -150 & \ch{Ti2Al3}            & $I4/mmm$          & -310 & -322 \\
\ch{Ti5Al2}            & $P4/mbm$              & -54  & -74  & \ch{Ti2Al3}            & $P4/mbm$          & -73  & -152 \\
\ch{Ti11Al5}           & $P4/mmm$              & -236 & -292 & \ch{Ti3Al5}            & $P4/mmm$          & -283 & -300 \\
\ch{Ti2Al} ($B8_{2}$)  & $P6_{3}/mmc$          & -306 & -304 & \ch{Ti3Al5}            & $P4/mmm$          & -158 & -195 \\
\ch{Ti2Al}             & $C2/m$                & -268 & -298 & \ch{TiAl2} ($C{14}$)  & $P6_{3}/mmc$       & -318 & -324 \\
Ti2Al ($C22$)          & $P321$                & -228 & -224 & \ch{TiAl2}             & $C2/m$            & -244 & -265 \\
Ti2Al($C40$)           & $P6_{2}22$            & -269 & -259 & \ch{TiAl2}             & $I4/mmm$          & -240 & -252 \\
\ch{Ti2Al}             & $ I4/mmm$             & -143 & -151 & \ch{TiAl2}($C38$)      & $I4/mmm$          & -231 & -232 \\
\ch{Ti2Al} ($C38$)     & $I4/mmm$              & -139 & -180 & \ch{TiAl2}             & $P4/mmm$          & -41  & -107 \\
\ch{Ti2Al}             & $Cmce$                & -49  & -59  & \ch{TiAl2}             & $I4_{1}/amd$      & -430 & -430 \\
\ch{Ti5Al3}            & $P\bar{6}2m$          & -116 & -108 & \ch{Ti2Al5}            & $P4/mbm$          & -290 & -308 \\
\ch{Ti3Al2}            & $P4/m$                & -332 & -341 & \ch{TiAl3} ($D0_{22}$) & $I4/mmm$          & -398 & -394 \\
\ch{Ti3Al2} ($D5_{a}$) & $P4/mbm$              & -245 & -241 & \ch{TiAl3} ($D0_{24}$) & $P6_{3}/mmc$      & -339 & -374 \\
\ch{Ti4Al3}            & $P6_{3}/m$            & -329 & -313 & \ch{TiAl3} ($D0_{19}$) & $P6_{3}/mmc$      & -320 & -327 \\
\ch{Ti4Al3}            & $I4/mmm$              & -217 & -223 & \ch{TiAl3} ($D0_{23}$) & $I4/mmm$          & -403 & -410 \\
\ch{TiAl}($L1_{0}$)    & $P4/mmm$              & -405 & -405 & \ch{TiAl4}($D1_{a}$)   & $I4/m$            & -217 & -271 \\
\ch{TiAl}              & $I4_{1}/amd$          & -368 & -369 & \ch{TiAl5}             & $P6_{3}2\bar{2}$  & -182 & -213 \\ \hline \hline
\end{tabular*}
}
\end{table}
\autoref{do19 props} shows that RANN is capable of modeling accurate material properties for the formed D0$_{19}$ structure.\\
\begin{table}[!htbp]
\caption{D0$_{19}$ properties :  \emph{italic} are from previous DFT. }
\label{do19 props}
\resizebox{\columnwidth}{!}{%
\begin{tabular*} {1.2\textwidth}{@{\extracolsep{\fill}}lcc}
\hline \hline
Properties        & Expt./DFT & RANN    \\ \hline
$a_0$ ($\AA$)             & 5.81$^{f}$/\emph{5.76$^{e}$},5.74 & 5.74    \\
$c_0$ ($\AA$)             & 4.65$^{f}$/\emph{4.66$^{e}$},4.64 & 4.64    \\
$c/a$ ratio       & 0.801$^{f}$/\emph{0.81$^{e}$},0.81 & 0.81    \\
$C_{11}$ ($GPa$)          & 183$^{a}$/\emph{221$^{b}$, 185.8$^{c}$, 184.7$^{d}$, 192.2$^{e}$},197.42 & 204.6   \\
$C_{12}$ ($GPa$)          & 89$^{a}$/\emph{71$^{b}$, 76.6$^{c}$, 82.4$^{d}$, 80.5$^{e}$},78.56 & 74.74   \\
$C_{13}$ ($GPa$)          & 63$^{a}$/\emph{85$^{b}$, 60.1$^{c}$, 63.4$^{d}$, 62.5$^{e}$},66.24 & 71.95   \\
$C_{33}$ ($GPa$)          & 225$^{a}$/\emph{238$^{b}$, 225.2$^{c}$, 225.1$^{d}$, 232.9$^{e}$},234.86 & 234.07  \\
$C_{44}$ ($GPa$)          & 64$^{a}$/\emph{69$^{b}$, 57.2$^{c}$, 54.0$^{d}$, 61.6$^{e}$},66.63 & 72.99   \\
$C_{66}$ ($GPa$)          & 47$^{a}$/\emph{75$^{b}$, 54.6$^{c}$, 51.2$^{d}$, 55.85$^{e}$},59.43 & 64.94   \\
$C_{12}-C_{66}$ ($GPa$)   & 42$^{a}$/\emph{-4$^{b}$, 22.0$^{c}$, 31.2$^{d}$, 24.65$^{e}$},19.03 & 9.8     \\
$C_{13}-C_{44}$ ($GPa$)   & -1$^{a}$/\emph{16$^{b}$, 2.9$^{c}$, 9.4$^{d}$, 0.9$^{e}$},-0.39 & -1.04   \\
$E_{\text{surf}}^{\hkl{0001}}$ ($\frac{mJ}{m^2}$) & /1.952 & 1.968   \\
$E_{\text{surf}}^{\hkl{1-100}}$ ($\frac{mJ}{m^2}$) & /2.313 & 2.254   \\
$E_{\text{surf}}^{\hkl{11-21}}$ ($\frac{mJ}{m^2}$) & /2.121 & 2.113   \\
$T_{\text{melt}}$ ($K$)        & & 2256.27 \\
\hline \hline
\multicolumn{3}{l}{\small $^{a}$\cite{tanaka1996elastic}, $^{b}$\cite{fu1990elastic}, $^{c}$\cite{liu2007first}, $^{d}$\cite{wei2010site}, $^{e}$\cite{zhang2017first}, $^{f}$\cite{pearson2013handbook} } \\
\end{tabular*}
}
\end{table}

\autoref{l10 props} shows that RANN is capable of reproducing near DFT results for the L1$_0$ material properties.
\begin{table}[!htbp]
\caption{L1$_{0}$ properties :  \emph{italic} are from previous DFT. }
\label{l10 props}
\resizebox{\columnwidth}{!}{%
\begin{tabular*} {1.2\textwidth}{@{\extracolsep{\fill}}lcc}
\hline \hline
Properties        & Expt./DFT & RANN    \\ \hline
$a_0$ ($\AA$)             & 3.997$^{a}$/\emph{3.996$^{f}$},3.98 & 3.98    \\
$c_0$ ($\AA$)             & 4.077$^{a}$/\emph{4.076$^{f}$},4.07 & 4.07    \\
$c/a$ ratio       & 1.02$^{a}$/\emph{1.02 $^{f}$},1.02 & 1.02    \\
$C_{11}$ ($GPa$)          & 187$^{b}$/\emph{190$^{c}$, 170$^{d}$, 164$^{e}$, 167.4$^{f}$},168.92 & 170.62   \\
$C_{12}$ ($GPa$)          &74.8$^{b}$/\emph{105$^{c}$, 88$^{d}$, 85.5$^{e}$, 87.7$^{f}$},87.2 & 92.93   \\
$C_{13}$ ($GPa$)          & 74.8$^{b}$/\emph{90$^{c}$, 84$^{d}$, 81$^{e}$, 86.1$^{f}$},86.93 & 87.37   \\
$C_{33}$ ($GPa$)          & 182$^{b}$/\emph{185$^{c}$, 179$^{d}$, 178.6$^{e}$, $^{f}$},169.14 & 176.21  \\
$C_{44}$ ($GPa$)          & 109$^{b}$/\emph{120$^{c}$, 115$^{d}$, 109.6$^{e}$, 111.1$^{f}$},110.33 & 102.54   \\
$C_{66}$ ($GPa$)          & 81.2$^{b}$/\emph{50$^{c}$, 72$^{d}$, 72.6$^{e}$, 73.7$^{f}$},70.15 & 65.09   \\
$C_{12}-C_{66}$ ($GPa$)   & -6.4$^{b}$/\emph{55$^{c}$, 16$^{d}$, 12.19$^{e}$, 14$^{f}$},22.78 & 27.84     \\
$C_{13}-C_{44}$ ($GPa$)   & -34.2$^{b}$/\emph{-30$^{c}$, -31$^{d}$, -28.6$^{e}$, -25$^{f}$},-22.96 & -15.18   \\
$E_{\text{surf}}^{\hkl{100}}$ ($\frac{mJ}{m^2}$) & /1.643 & 1.798   \\
$E_{\text{surf}}^{\hkl{111}}$ ($\frac{mJ}{m^2}$) & /1.667 & 1.714   \\
$T_{\text{melt}}$ ($K$)        & & 2279.11 \\
\hline \hline
\multicolumn{3}{l}{\small $^{a}$\cite{pei2021systematic}, $^{b}$\cite{tanaka1996single}, $^{c}$\cite{fu1990elastic}, $^{d}$\cite{liu2007first}, $^{e}$\cite{fu2010ab}, $^{f}$\cite{zhang2017first} } \\
\end{tabular*}
}
\end{table}

\begin{figure}[!htbp]
\noindent \centering
\subfloat{\includegraphics[width=0.75\textwidth]{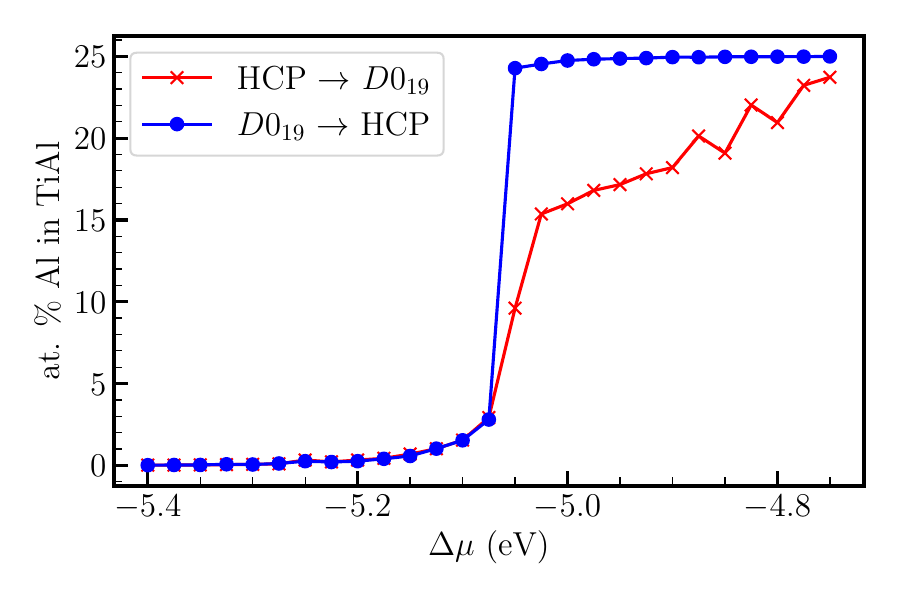}}
\caption{Semi-Grand-Canonical Monte Carlo Simulations for MTP \cite{qi2023machine} at 900 K.}
\label{mc MTP}
\end{figure}

The $\alpha$ to D0$_{19}$ transition was examined with the MTP \cite{qi2023machine} potential using the semi-grand-canonical Monte Carlo (MC) method \cite{sadigh2012scalable} as done in the work of \citet{kim2016atomistic}. This method adds a chemical potential difference term, $\Delta\mu$, to the Metropolis acceptance criterion \cite{metropolis1953equation} in order to determine if it is acceptable to swap one type of atom with another. Pure $\alpha$-Ti structures and D0$_{19}$ structures consisting of 1024 atoms were put under the semi-grand-canonical MC fix in LAMMPS at varying temperatures and $\Delta\mu$ values. A hysteresis loop is found by plotting the equilibrated atomic percentage of Al against the $\Delta\mu$ values. The transition point for a given temperature is taken to be the atomic percentage of Al that corresponds to the final data point before the original phase transforms. The MTP potential displays an accurate transition from the D0$_{19}$ phase to the $\alpha$ phase; however, it finds an intermediate phase going from the $\alpha$ phase to the D0$_{19}$ phase. This is evident in the hysteresis loop shown in \autoref{mc MTP} as the D0$_{19}$ phase jumps from roughly 25\% Al to the $\alpha$ phase at roughly 3\% Al, but the $\alpha$ phase does not directly jump to 25\% Al. The $\alpha$ phase finds a stable structure around 15\% Al before slowly reaching the D0$_{19}$ phase.
\clearpage
\newpage
\bibliography{TiAl_ref}